\documentstyle[preprint,prb,aps,12pt]{revtex}
\long\def\title#1{{\Large\begin{center}#1\end{center}\par}}
\long\def\address#1{\begin{center}#1\end{center}\par}
\long\def\author#1{\begin{center}#1\end{center}\par}

\tightenlines
\begin{document}

\draft

\title{Inter-layer slide and stress relaxation in a bilayer fluid 
membrane in the patch-clamp setting.}

\author{Sergei I. Mukhin and Svetlana V. Baoukina}
\address{Theoretical Physics Department,
Moscow Institute for Steel and Alloys, Leninskii pr. 4, 119991 Moscow,
Russia}

\vspace{3mm}Running title: Inter-layer slide and stress relaxation

\vspace{3mm} Keywords: bilayer lipid membrane, mechanosensation,
lateral stress relaxation, interlayer slide, MscL gating, thermal activation
\vspace{3mm}
\maketitle
\begin{abstract}
Protein mechanosensitive channels (MS) are activated
 by tension  
transmitted through the lipid bilayer. We propose a 
theory of 
lateral stress relaxation in a bilayer lipid membrane 
exposed to external 
pressure pulse 
in the patch-clamp experimental setting. It is shown that
 transfer of lipid 
molecules into a strained region is thermodynamically 
advantageous due to local 
decrease of the stress. Considered stress relaxation mechanism
 may explain recent 
experimental observations (Davidson and Martinac 2003) of 
adaptation of MscL, bacterial mechanosensitive channel 
of large conductance, to sustained membrane stretch. 
Lateral stress relaxation in the monolayer, which controls the 
gating of 
MscL, triggers thermally activated transition of the open 
channels back to the closed state ("adaptation"). We evaluate 
the contribution of 
the hydrophobic mismatch between MS channel and lipid bilayer to 
the energy barrier separating open and closed states. Then, using the MscL 
thermodynamic model (Sukharev et. al, 1999), we estimate characteristic  
adaptation times at room temperature to be of the order of seconds, well 
in the range 
of the experimental data (Davidson and Martinac 2003). 
Estimated propagation time of the initial channel-opening stress  
over the whole membrane is 4-5 orders of magnitude shorter. 
\end{abstract}

\section{Introduction}
The functioning of the protein channels in the cell membranes regulates 
flows of ions in- and out of the cell, thus influencing signal transmission in 
the neural networks (Doyle et al., 1998; Ming Zhou et al. 2001). Mechanical 
stresses in the cell membranes control gating of the protein mechanosensitive
(MS) channels (Sukharev, 2001), which hence play a role of 
mechanoreceptors in the cellular organisms. Therefore, a theory of the stress
propagation and relaxation in the lipid bilayer membranes is of substantial interest 
for the fundamental and practical purposes (Yeung and Evans, 1995; Cantor, 1999).
 One of the biological objects convenient for experimental study
 is the large conductance MS channel (MscL) in the inner membrane
 of the Escherichia coli (E-coli) bacteria. It is possible to measure
 a single channel conductance, an find its dependence on
 the internal lateral tension in the membrane. The tension, which causes
 opening of the channel, could be evaluated using video microscopy measurements 
 of the membrane curvature formed under the applied external pressure 
(Sukharev et al., 1999).

 The time-dependence of the membrane's conductance observed in (Hase et al., 1995) 
 and (Davidson and Martinac, 2003) reveals gradual collapse of the 
 ionic currents through the MscL's (called adaptation) within seconds after their opening. 
 According to existing hypothesis (Sachs and Morris, 1998), this 
 may happen due to mutual slide of the lipid monolayers, constituting a
 bilayer membrane of the E-coli. This slide would then cause a relaxation of the 
 channel-opening lateral stress inside the membrane. Thus, the slide of the monolayers 
 is induced by the same pressure pulse, which leads initially to the opening of the MscL's.

In this paper we propose a theory of such a stress-relaxation process and demonstrate that 
the slide of the monolayers after application of a constant external pressure gradient across
the membrane is indeed thermodynamically advantageous.
In Section II we outline the main physical ideas and mechanisms considered in this paper.
In particular, we apply the theory of the interlayer slide dynamics (Yeung and Evans, 1995) 
in order to evaluate characteristic time of the stress-relaxation process. We also show
that characteristic time of the MscL adaptation following the stress-relaxation in the
membrane is actually much longer. In general, this is caused by the necessity of a thermal 
activation involved in the channel adaptation process. Using the MscL thermodynamic model 
(Sukharev et. al, 1999), we estimate characteristic adaptation times at room temperature 
to be of the order of a few seconds, well in the range of the recently reported experimental 
data (Davidson and Martinac 2003).

In Section III we introduce a model free energy functional of a bilayer membrane. 
We add to the free energy an adhesion energy term, which reflects important feature
of the patch-clamp experimental setting regarding adhesion of the lipid monolayer to
the glass wall of the pipette. Then, the self-consistent calculation procedure is outlined
of the membrane free energy in the semi- and complete equilibrium states under a 
constant pressure gradient in a patch-clamp setting. 
In Section IV we consider results of our numerical calculation of the free energy difference 
between the two successive semi- and complete equilibrium 
conformations of a membrane, i.e. : (1) after application of a constant external pressure 
gradient, but before the interlayer slide; and (2) after the slide of one monolayer with 
respect to the other, leading to the complete equilibrium conformation. The influence of the
adhesion of the lipids to the glass wall of the pipette is also studied numerically. Results
reasonably compare with the experimental data. \\ 
In Section V we review our main estimates of the time scales characterizing the membrane 
and the MscL dynamics.
Also, the effect of the energy barrier separating open and closed states of the MscL
on its adaptation to sustained stress is discussed. The difference between MscL adaptation 
in the thick and thin lipid bilayers is considered in relation with the 
experimental data for the PC-20 and PC-18 liposome patches described in the companion
paper (Davidson and Martinac 2003).\\
Possible experimental verifications and future improvements of the theory are also discussed.
We focus on the physical ideas in the main text and place more detailed mathematical
derivations in the appendices.

\section{Bilayer lipid membrane with MscL: characteristic time scales}
\subsection{The shortest time}
The first, semi-equilibrium conformation of a bilayer in the patch-clamp setting 
is achieved shortly after application of a pressure difference pulse $P$, within 
the time of propagation of the mechanical stress along a membrane: 
$\Delta t \sim \omega_b^{-1}\sim 10^{-5}\div 10^{-4}$sec. Here characteristic 
bending frequency $\omega_b$ is evaluated below in Eqs. (\ref{omegab}) and (\ref{deltat})
using the theory of bending waves in a thin plate. In the semi-equilibrium  state under
consideration the number of the phospholipid molecules in the curved part of the 
bilayer membrane has increased relative to the initial flat conformation mainly due to a 
tearing off the membrane from the walls of the pipette, see Fig. 1. 
In order to allow for this, a finite 
adhesion energy, $E_{ad}$, is introduced below. The tensions $T_1$ and $T_2$ in the 
two monolayers of the membrane approximately coincide, since the radius $R$ of the 
membrane is of the order of $1\div 5 \mu m$, while its thickness, $2\cdot l_0$, is about 
(Hase et al., 1995) $3.5\div 4 nm$. 
Hence, on such a short time scale $\Delta t$ the membrane behaves effectively 
like a unit structure, as if the monolayers are strongly coupled together and 
can not slide with respect to one another.
\subsection{The lateral stress relaxation time}
Described conformation is a semi-equilibrium one because it possesses a tension 
gradient along the surface of the lower monolayer $1$. The gradient $\nabla_s T_1$ 
arises
as long as the tension in the curved part of the membrane (in the pipette) is 
finite: $T_1\neq 0$, while it is zero in the rest of the patch, see Fig. 1. Hence, layer 
$1$ will slide against layer $2$ (the latter is stuck to the walls).
A sliding velocity $v_s$ is determined by the drag coefficient (Yeung and 
Evans, 1995) $b$ :

\begin{equation}
 bv_s=\nabla_s T_1={K_A}\nabla_s\alpha\,,
 \label{drag}
 \end{equation}

\noindent where $K_A$ and $\alpha$ are area compressibility modulus for monolayer 
 and dilation field respectively($\alpha\equiv a/a_0-1$ where $a$ is the area per lipid 
molecule). Then, a conservation equation for the dilation field takes the form of a
diffusion equation:

\begin{equation}
 \frac{\partial \alpha}{\partial t}=\nabla_s v_s\equiv {K_A}{\nabla_s}^2\alpha
 \equiv D{\nabla_s}^2\alpha\;\;;\; D=\frac{K_A}{b}\,,
 \label{difs}
 \end{equation}

 \noindent where $D$ plays the role of a diffusivity constant of the dilation field.
Then, the lateral stress relaxation time, $\tau_s $, which is necessary for the dilation 
field to level off diffusively between the outer patch and the center of the curved membrane 
along the distance $\sim R$ is evaluated as:

\begin{equation}
\tau_s \sim \displaystyle\frac{R^2}{D}\sim \frac{R^2b}{K}\sim 
\frac{2R^2b\bar{\alpha}}{T_1}\sim \frac{1.2\cdot 10^{-7}cm^2\cdot 10^7dyn\cdot 
sec/cm^3\cdot 0.05}
{6dyn/cm}=0.01 sec\,,
 \label{time}
 \end{equation}

\noindent where we evaluate $K_A\sim T_1/\bar{\alpha}$. The characteristic area dilation per 
lipid, which does not destroy the membrane, is evaluated as $\bar{\alpha}\leq 0.05$. 
A typical value of $b$ for a flat membrane is (Yeung and Evans, 1995) 
$\sim 10^7 dyn\cdot sec/cm^3$. 
Measured value of external pressure (Sukharev et al., 1999) is about $50 mm$Hg, 
which corresponds to the membrane's lateral tension $T=11.8 dyn/cm$ at the 
curvature 
radius $r\sim R\approx 3.5 \mu m$. Hence, the instantaneous tension in the 
monolayer $1$ is: $T_1=T/2\sim 6 dyn/cm$, thus leading to above estimate of the 
lateral stress relaxation time $\tau_s$.
\par The time $\tau_s$ may 
become even greater if interdigitation between the monolayers increases in the 
curved-stretched conformation of the membrane. The effective drag coefficient $b$ may 
also be enhanced due to an extra impedance caused by a finite density of the 
protein channels piercing the membrane.
\par The estimate in 
Eq. \ref{time} is still considerably shorter than the experimental 
adaptation times $\sim 1 sec$ (Hase et al., 1995), (Davidson and Martinac 2003).
In order to understand the reason for such a difference one has to allow for additional factor 
influencing the MscL adaptation time. After the relaxation of the lateral tension $T_1$ 
in the lower monolayer the MscL has to overcome a (free)energy barrier to make transition
from the open to the closed state. This transition, as is estimated below, requires time, 
which, indeed, proves to be much longer than $\tau_s$.  

\subsection{The MscL adaptation time}

The structure of MscL is assymmetric and the cytoplasmic half of the 
channel poses a barrier to ion permeation (Sukharev et al., 2001). 
This half of the channel is situated in the lower monolayer of the membrane in the 
considered experimental settings (Sukharev et al., 1999; Hase et al., 1995). 
Hence, activation/closing of the MscL is regulated by the lateral tension $T_1$ in this 
lower monolayer $1$. Let us use a simple elastic model of MscL channel (Sukharev, Sigurdson
et al., 1999) to study the kinetics of the channel closing induced by interlayer slide. 
A detailed derivation is presented in the Appendix B (see also Fig. 4). Here we merely 
describe the main consequences.
The value of the rate constant for the open to closed transition, $k_c$, is the key factor:
\begin{equation}
k_c=\displaystyle\tilde{k}\exp{\{-E_{act,c}/k_BT_0\}}\;,
\label{rate}
\end{equation}

\noindent where $k_B$ is Boltzmann'constant, $T_0$ is temperature, $\tilde{k}$ is 
"attempt rate". 
The activation barrier for closing of the channel starting from the open state equals (see
Appendix B, Eq. (\ref{barc})):
\begin{equation}
 E_{act,c}(T)=\frac{B_0}{2}(A_0-A_b)^2+T(A_0-A_b)+\frac{T^2}{2B_0}
\label{barrierc}
\end{equation}

\noindent where $T$ is lateral tension stretching the channel; $A_{0}>A_{b}$ designate the 
areas of the pore in the open and top-of the -barrier conformations respectively; 
$B_0$ is elastic constant of the channel in the open state.
The parabola Eq. (\ref{barrierc}) has minimum at $T<0$. Hence, the activation barrier for 
the channel closing remains finite (does not reach zero) within the interval of non-negative 
tensions ($T>0$). Simultaneously, we find that relaxation of the tension $T_1$ in the lower 
monolayer favours closing of the MscL with the rate $k_c$ defined in Eq. (\ref{rate}), 
as $E_{act,c}$ in (\ref{barrierc}) decreases with the decrease of $T$. The pre-exponential 
factor $\tilde{k}$ in Eq. (\ref{barrierc}) may be estimated as $\sim 100 sec^{-1}$ 
(Sukharev and Markin, 2001). Then, regarding adaptation time as the equilibration time of 
the initially open channel $\tau_{eqv}=k_c^{-1}\sim 1.5 sec$ (Davidson and Martinac, 2003)), 
triggered by the relaxation of lateral stress in 
the lower monolayer, we estimate the order of magnitude of the activation energy $E_{act,c}$
inverting Eq. (\ref{rate}):

\begin{equation}
E_{act,c}=k_BT_0\ln{\tilde{k}/k_c}\equiv k_BT_0\ln{\{\tilde{k}\tau_{eqv}\}}\approx
k_BT_0\ln\{150\}\approx 5k_BT_0\,.
\label{eact_c}
 \end{equation}

\noindent We show in Section V that this value of $E_{act,c}$ has the relevant scale for the
known set of the thermodynamic parameters characterizing the MscL channels.

\subsection{The MscL activation time}
The necessity to overcome a free energy barrier may also increase MscL transition time from 
the closed to the open state (MscL activation time). Remarkably, unlike in the case of 
adaptation, the hight of the barrier for MscL activation may vanish at finite tension $T^*>0$.
Thus, the lateral tension $T$ enhances MscL opening rate, and the latter process may happen 
without the thermal activation. 
Using the same elastic model as in Eqs. (\ref{rate})-(\ref{barrierc}) above, we find expression 
for the rate constant of the closed to open state transition, $k_o$ :

\begin{equation}
k_o=\displaystyle\tilde{k}\exp{\{-E_{act,o}/k_BT_0\}}\;,
\label{rateo}
\end{equation}

\noindent where $\tilde{k}$ is "attempt rate" in the closed state.
 The activation barrier for opening of the channel starting from the closed state equals 
(see Appendix B, Eq. (\ref{baro})):

\begin{equation}
 E_{act,o}(T)=\frac{B_0}{2}(A_b-A_C)^2-T(A_b-A_C)+\frac{T^2}{2B_C}\;,
\label{barriero}
\end{equation}

\noindent where $A_{b}$ designates the area of the pore in the closed MscL conformation; 
$B_C$ is elastic constant of the channel in the closed state. The maximal value 
$E_{act,o}(T)$ possesses at $T=0$:

 \begin{equation}
 E_{act,o}(T=0)=\frac{B_0}{2}(A_b-A_C)^2\;,
\label{barriero1}
\end{equation}
 
\noindent see Fig. 4.
The minimal value 
$E_{act,o}(T)=0$ is reached at $T=T^*$, where :

\begin{equation}
 T^*=B_C(A_b-A_C)\,;\;\; E_{act,o}(T\geq T^*)=0\,.
\label{tstar}
\end{equation}

\noindent Hence, it follows from Eqs. (\ref{rateo}) and (\ref{barriero1}), (\ref{tstar}) 
that activation time $\tau_a\sim k_o^{-1}$ changes from its maximal value $\tau_a(T=0)$ at 
zero tension down to its shortest value at $T\geq T^*$:
\begin{eqnarray}
\tau_a(T)=\left\{\begin{array}{l}
\tilde{k}^{-1}\exp{\{{B_0}(A_b-A_C)^2/2k_BT_0\}}\;, T=0\nonumber\\
\tilde{k}^{-1}\;, T\geq T^*  
\end{array}\right.
\label{tau_a}
\end{eqnarray}

\noindent Indeed, according to (Sukharev, Sigurdson et al., 1999), an increase of $T$ from
$11.9mN/m$ to $14.14 mN/m$ leads to a decrease of the activation time $\tau_a\sim k_o^{-1}$
by about 40 times, so that it reaches the absolute value of $6.8\cdot 10^{-3}sec$.
\subsection{The lipid diffusion time}
We have also calculated the free energy of the whole patch: bent part and unstrained
outer part, as a function of the external pressure difference $P$ in both 
the semi- and complete equilibrium states.
Indeed, the free energy in the relaxed, $T_1=0$ state, proves
to be lower than in the semi-equilibrium state achieved before the slide,
Fig. 2{\it{a}}. A comparison of the calculated numbers $N_1$ and $N_2$ of the lipids, 
in the 
curved lower and upper monolayers correspondingly, in the semi- and complete 
equilibrium states follows from Fig. 2{\it{b}}. It shows that the number 
$\Delta N_1$ 
of the lipids "sucked in" from the patch to relax $T_1$ is of a macroscopic 
magnitude, e.g. $\Delta N_1\sim 4 {\%} N_1$. Based on these result, we estimate 
also a diffusion time, $\tau_d$, which would be necessary for the lipid molecules 
to move from the patch to the monolayer $1$ chaotically, i.e. without the slide 
motion of the monolayer as a whole. 
Experimentally determined diffusion coefficient of the individual lipid molecules 
in bio-membranes at room temperature is (Sonnleitner et al., 1999): 
$D\sim 1\div 10 \mu m^2/sec$. 
Hence, we evaluate a characteristic diffusion time $\tau_d$ as:

\begin{equation}
\tau_d \sim \displaystyle (\Delta N_1/N_1)R^2/D \sim 0.1\div 1
\mbox{sec},
 \label{diffusion}
 \end{equation}

\noindent which is by orders of magnitude longer than the sliding time 
$\tau_s$, Eq. \ref{time}. Hence, diffusion of individual lipid molecules proves  
to be less effective for the relaxation of the stress in the concave 
monolayer than the interlayer slide.

\subsection{Adaptation in PC-20 {\it{versus}} no adaptation in PC-18}
Despite the seeming clarity of the mechanisms, which determine the different time scales
described above, the whole picture of MscL response to the sustained lateral tension in
the membrane can be more involved: hydrophobic mismatch affects MscL gating (Perozo et al.,
2002). While MscL exhibited adaptation in bilayer formed with diecosenoyl
phosphatidycholine, PC-20 molecules, no adaptation to sustained stress 
was observed in bilayer made of dioleoyl phosphatidycholine, PC-18 
(Davidson and Martinac 2003).
Below we present some estimates demonstrating that the change of MscL hydrophobic mismatch 
between its closed and open conformations can be responsible for a fairly different gating
behavior depending on whether the lipid bilayer is "thick" or "thin". This may explain
experimentally observed drastic difference betwee MscL gating behaviors in PC-20 and PC-18 
bilayers. We use here again the elastic model of MscL channel (Sukharev and Mrkin, 2001) 
in combination with the lipid-protein hydrophobic mismatch model (Fournier, 1999) shortly 
introduced in the Appendix.\\

The length of hydrophobic region of (the external wall) MscL channel is different for its
closed and open conformations (Sukharev et al., 2001).  
The mismatch energies in the open and closed states of the MscL channel, $F_{LP}^{O}$ and, 
$F_{LP}^{C}$, contribute to the total free energy of these conformations, $F_O$ and $F_{C}$.
The free energy difference between open and closed states is then given by:

\begin{equation}
F_O-F_C=\tilde{E}_0+\Delta F_{LP}\,,
\label{lp}
 \end{equation}

\noindent
here $\Delta F_{LP}=F^O_{LP}-F^C_{LP}$ is the part of the energy difference arising from the 
difference in 
hydrophobic mismatch energies, while $\tilde{E}_0$ is determined by other factors
(Sukharev, Sigurdson et al., 1999), e.g. the type of lipids in the bilayer (hydrocarbon 
chains lengh; curvature stress profile), thermodynamic conditions of the surrounding 
solution (pH, temperature). In the elastic model of MscL at $T=0$: 
$\tilde{E}_0+\Delta F_{LP}=E_0$.
The mismatch energy depends on the type of lipid; if the length of the lipid tail provides
 better matching of lipid-protein hydrophobic regions, this results in lower mismatch energy 
and lower total energy of the given channel conformation. In other words, lipids surrounding 
the channel favour certain (closed or open) conformation of the MscL 
(Hamill and Martinac, 2001).

For example, if in the (initial) closed state of MscL the hydrophobic thickness of the bilayer
is equal or greater then that of the channel, the channel opening increases the mismatch and, 
thus, the mismatch energy adds to the value $E_0$ (and to the energy separation $F_O-F_C)$.
On the other hand, if the hydrophobic thickness of the bilayer is less or matches that of MscL
in the open conformation, the channel opening decreases the mismatch and, hence, the
mismatch energy subtracts from $E_0$. In the thermodynamic model of MscL it corresponds, 
see Fig. 4, to the decreased vertical distance (i.e. $E_0$ at $T=0$) between the parabolas 
describing the free energies of the open and closed conformations of the channel. When MscL 
opens, its hydrophobic thickness decreases by $6\div 8\AA$ (S. Sukharev, Maryland University,
personal communication, 2003). Our estimate (see Appendix B) shows, that the difference of 
mismatch energies between open and closed states of MscL, $\Delta F_{LP}$, may range up 
to $20k_BT$.

We would expect that shorter lipids (e.g. PC-18 compared to PC-20) favour open state of 
MscL:

\begin{equation}
(\Delta F_{LP})^{PC18}<(\Delta F_{LP})^{PC20}\,.
\label{dlp}
 \end{equation}

\noindent
This implies that the energy difference, $F_O-F_C$, becomes smaller:

\begin{equation}
(F_{O}-F_C)^{PC18}<(F_{O}-F_C)^{PC20}\,.
\label{dflp}
 \end{equation}

\noindent
Thus at any tension the ratio of the equilibrium fractions of the open and closed 
channels, $n_O/n_C$ : 

\begin{equation}
n_{O}/n_C\sim \exp{\{-(F_O-F_C)/k_BT_0\}}
\label{dn}
 \end{equation}

\noindent
is shifted towards increased fraction of open channels $n_O$ for PC-18 as compared to PC-20 :

\begin{equation}
(n_{O}/n_C)^{PC18}>(n_{O}/n_C)^{PC20}\,.
\label{pc1820}
 \end{equation}

\noindent
This is why in PC-18 liposome an interlayer slide (and resulting relaxation of the lateral 
tension in the lower monolayer) may not lead to a complete closing back of the MscL's in the
final thermodynamic equilibrium state, since at $F_O\approx F_C$ we have $n_O/n_C\sim 1$. 
We also expect the closing of (a fraction of ) MscL's to take longer time in PC-18 bilayer 
in comparison with PC-20 bilayer as the energy barrier between closed and open conformations,
$E_{act,c}$,
increases with a decrease of $E_0$, Fig. 5 (see explanation 
in Appendix B after Eq. (\ref{baro})). 
The higher energy barrier leads to the lower rate of closed-to-open transition, 
$k_c$ in Eq.(\ref{rate}).

\subsection{Interlayer slide and hydrophobic mismatch}
Interlayer slide makes no significant contribution to the energy of hydrophobic mismatch.
Due to the incompressibility condition,$V=a\cdot l=const$ ($V$ - is volume of lipid molecule, 
$a$ - is area per lipid molecule, $l$-is lipid length in the strained membrane), tension in 
each monolayer, $T_i$, produces fractional change in its thickness:

\begin{equation}
T_i=-\frac{\Delta l}{l_0}\cdot K_A\,.
\label{tensions}
 \end{equation}

\noindent
Here $i=1,2$ is monolayer label;
$\Delta l=l-l_0$, and $l_0$ is the length of a lipid in the unstressed flat membrane;
$K_A$ is the area compressibility (expansion) modulus of a monolayer.
After the application of external pressure, $P$, to the membrane (but before interlayer slide)
both monolayers are strained and tensions in them approximately coincide:

\begin{equation}
T_1\approx T_2\approx T/2\,,
\label{tensions1}
 \end{equation}

\noindent
here $T$ is tension in the membrane defined according to Laplace law: $T=Pr/2$, 
$r$ is curvature radius.
The total relative change in membrane thickness, is given by:

\begin{equation}
\frac{\Delta l_1+\Delta l_2}{2l_0}\approx \displaystyle \left[-\frac{T_1}{2K_A}\right]
+\left[-\frac{T_2}{2K_A}\right]=-\frac{T}{2K_A}\,.
\label{thickness}
 \end{equation}

\noindent
After the interlayer slide the flow of lipids from the unstrained outer patch to the 
lower monolayer leads to the relaxation of lateral stress in it; tensions are redistributed:

\begin{equation}
T_1=0\,,\; T_2=T\,.
\label{ties}
 \end{equation}

\noindent
But the total change in the bilayer thickness, evidently, remains nearly the same:

\begin{equation}
\Delta l_1=0\,,\; \frac{\Delta l_2}{2l_0}\approx -\frac{T}{2K_A}\,.
\label{thickness1}
 \end{equation}

\section{The model free energy of membrane}

We start from the microscopic model of a bilayer membrane as described
e.g. in (Safran, 1994; Ben-Shaul, 1995; Xiang and Anderson, 1994). First, write 
a single i-th monolayer
($i=1,2$) free energy per lipid molecule $f_i$:

\begin{eqnarray}
f_i=f_{si}+f_{hi}+f_{ti}\equiv \gamma a_i+{C}/{a_i}+
\displaystyle({k_{s}}/{2})(l_i-l_s)^2\label{fren}
\end{eqnarray}

\noindent Here $f_{si}$ is the external surface energy of the
monolayer, which increases with the area per lipid molecule $a_i$
due to the hydrocarbon tails interaction energy with the solvent.
The short-range repulsion between the polar molecular heads at the
surface of the membrane is represented by the second term in Eq.
\ref{fren}. The third term, though looks like a spring elastic
energy (with not stretched spring length $l_s$), codes for the
entropic  repulsion between the hydrocarbon tails in the depth of
each monolayer (Safran, 1994). 
In what follows we omit the second term in Eq. \ref{fren} on the empirical 
grounds (Ben-Shaul, 1995) relevant for
the chain molecules in phospholipid bilayers. There is no
interface energy term included in Eq. \ref{fren} for the internal
surfaces of the monolayers forming the bilayer, as the latter are
not reachable for the solvent molecules. We also assume a
vanishing interdigitation of the tails between the adjoint
monolayers. The length of a lipid molecule $l_i$, i.e. the
thickness of the i-th monolayer, is not an independent variable
from the aria per molecule $a_i$. The volume $v$ per lipid
molecule is conserved, i.e. the latter could be considered as
incompressible (Safran, 1994; Ben-Shaul, 1995). Due to this conservation 
condition
the length $l_i$ is related to the mean and gaussian curvatures
$H$ and $K$ at the interface between the monolayers by a well
known differential geometry formulas (Safran, 1994) :

\begin{eqnarray}
&& l_1=l_{01}-l_{01}^2H+\displaystyle({2}/{3}) l_{01}^3K
\label{curvs1}\\
 && l_2=l_{02}+l_{02}^2H+\displaystyle({2}/{3})
l_{02}^3K \label{curvs2}
\end{eqnarray}

\noindent Here the change of sign in front of $H$-term in the
equation Eq. \ref{curvs2} relative to that in equation
Eq. \ref{curvs1} is due to a simple geometrical fact that the {\em
external} normal vector to e.g. the layer 1 is simultaneously an
{\em internal} normal vector to the layer 2 at their mutual
interface. Different $l_{0i}$ parameters just reflect the
incompressibility of the lipid molecules mentioned above:

\begin{eqnarray}
l_{0i}=v/a_i \label{lis}
\end{eqnarray}

\noindent Substituting  Eqs. \ref{curvs1} and \ref{curvs2} into
the third term of Eq. \ref{fren} one finds contributions to the
free energy from the tails in the form:

\begin{eqnarray}
&&f_{t1}=({k_sl_{01}^4}/{2})\left[(H-C_{01})^2+
({4}/{3})C_{01}l_{01}K\right]
\label{ftis1}\\
&&f_{t2}=({k_sl_{02}^4}/{2})\left[(H+C_{02})^2+
({4}/{3})C_{02}l_{02}K\right]
\label{ftis2}
\end{eqnarray}

\noindent where parameters $C_{0i}$ have the meaning of the local
spontaneous curvatures of the $i$-th layer:

\begin{eqnarray}
C_{0i}=(l_{0i}-l_s)/l_{0i}^2 \label{cis}
\end{eqnarray}

We readily recognize the Helfrisch's  formula (Zhong-Can and Helfrich, 1987;
Zhong-Can and Helfrich, 1989) for the
free energy of a membrane in Eqs. \ref{ftis1} and \ref{ftis2}.


In a symmetric bilayer case the numbers of lipids in the 1-st and 2-nd
monolayer are equal: $N_1=N_2$. 
Under the equivalent conditions on the
opposite surfaces of the membrane: $a_1=a_2$, the linear in $H$
terms in Eqs. \ref{ftis1} and \ref{ftis2} would cancel in the
total free energy $F$:

\begin{eqnarray}
F=\sum_{i=1,2}F_i=\sum_{i=1,2}N_if_i \label{frent}
\end{eqnarray}

\noindent Nevertheless, as shown below, a linear in $H$ term may
arise when the up {\it vs} down symmetry of a membrane is broken,
e.g. by applied pressure gradient in combination with the
different boundary conditions at the monolayers peripheries, see
Fig. 1.

To make the whole idea transparent we restrict our present
derivation to the case of a "spherically homogeneous
distributions" of molecules, i.e. considering $a_i$'s as being
different for the different indices i's, but position-independent
within a curved part of the i-th monolayer. Below, we neglect the 
inhomogeneous distribution of the strain across the thickness of each 
monolayer.
Also, we consider only spherical shapes of the curved part of the 
bilayer membrane, while introducing position independent (over the
membrane's surface) mean and gaussian curvatures:

\begin{eqnarray}
H=1/r\,;\;\;\; K=H^2=\displaystyle 1/r^2
\label{curvs}
\end{eqnarray}

\noindent where $r$ is the radius of curvature. The radius of the
base of the curved membrane is fixed at $R\leq r$, in accord with
the fixed radius of the pipette, which sucks in the membrane, and
creates a pressure difference $P$ between inside and outside
surfaces of the membrane in the experimental setup 
(Sukharev et al., 1999; Hase et al., 1995).
Then, according to the Laplace's law we have:

\begin{equation}
T_1+T_2=Pr/2\,,
\label{laplace}
\end{equation}

\noindent where $T_i$ is the lateral tension in the i-th monolayer, and we 
neglected a small difference between the curvatures of the monolayers.

First, we consider the free energy of a flat, undeformed bilayer membrane, in 
order to use it as a reference value of the free energy per lipid molecule, 
$f_0$. 
The flat membrane has a monolayer thickness $l_0=v/a_0$, where the area per 
molecule $a_0$ is determined from the condition of the minimum of the free 
energy $F$ defined in Eq. \ref{frent}:

\begin{eqnarray}
&\displaystyle\partial F/\partial a_0=0\;,\;\;\;
\mbox{at:} \label{mini}\\
& a_1=a_2=a_0,\;\;\; H=K=0,\;\;\;
N_1=N_2=N_0.
\label{a0}
\end{eqnarray}

\noindent Using conditions from Eq. \ref{a0} in the Eqs. \ref{fren} - 
\ref{frent} we derive the following expression for the free energy per lipid
molecule, $f_0$, in the undeformed state: 

\begin{eqnarray}
f_0=f_1(z_0,h=0)=f_2(z_0,h=0)=\epsilon_sz_0+
\epsilon_t\{1+z_0^{-2}-2z_0^{-1}\}\,, \label{fredimless2}
\end{eqnarray} 

\noindent where the following dimensionless parameters have been introduced:

\begin{eqnarray}
&&\displaystyle z_i={a_il_s}/{v};\;\;h={l_s}/{r}\equiv
Hl_s;\;\;h_0={l_s}/{R};\;\;\epsilon_s={\gamma
v}/{l_s};\;\;\epsilon_t={k_sl_s^2}/{2}.\label{dimless}
\end{eqnarray}

\noindent
It is useful for the comprehension of the rest of
the paper to mention here orders of magnitude of the main
parameters. Using the values of $R$ and $l_s\sim l_0$ mentioned
in the Introduction, we find that the dimensionless curvature 
is of the order: $h\sim 10^{-3}\ll 1$. Therefore, it could be
inferred from  Eqs. \ref{curvs1} and \ref{curvs2}, and from 
the definition Eq. \ref{lis}, that $v\approx l_s a_i $, which
leads to an estimate: $z_i\sim 1$. The entropic nature of the
$f_{ti}$ term in Eq. \ref{fren} is reflected in the temperature
dependent coefficient $k_s$, which for the known phospholipid
bilayers is of the order (Ben-Shaul, 1995; Xiang and Anderson, 1994): 
$k_sl_s^2\sim 25 kT$,
where $k$ is the Boltzman's constant and $T$ is the absolute
temperature. The scale of the surface term $f_{si}$ in Eq.
\ref{fren} is characterized by the coefficient $\gamma$, which
at room temperature is of the order (Ben-Shaul, 1995): $\gamma\sim 0.1
kT/\AA^2$. Gathering all the estimates, and also taking into
account that the typical value of an area per molecule is (Ben-Shaul, 1995): 
$a_i\sim 60\AA^2$, we obtain the following list of the
estimates (taken for the room temperature $T\sim 300K$):

\begin{eqnarray}
&&\displaystyle z_i\sim 1;\;\;h\sim 10^{-3};\;\;\epsilon_s\sim
3\cdot 10^{-13}\mbox{erg};\;\; \epsilon_t\sim 1.3\epsilon_s.
\label{dimvalues}
\end{eqnarray}

\noindent An additional important dimensionless parameter relates
characteristic experimental value of the pressure difference $P
\sim 50$mm Hg (Sukharev et al., 1999; Hase et al., 1995) with the 
"microscopical bending energy" $\epsilon_t$:

\begin{eqnarray}
\displaystyle P v/(2\epsilon_t)\sim 
10^{-4}\;\;\;\mbox{at:}\;\;\epsilon_t\sim 5\cdot 10^{-13}\mbox{erg};\;\;
v\sim 1.2\cdot 10^{3}\AA^{3} , \label{dimp}
\end{eqnarray}

\noindent where we use an estimate $l_s\sim 20\AA$ for the typical
length of a free phospholipid molecule (Ben-Shaul, 1995) and 
$\epsilon_s=0.6\epsilon_t$.

\noindent
In the dimensionless parameters and variables equation Eq. \ref{mini} reads:

\begin{eqnarray}
\displaystyle\partial{f_0}/\partial{z_0}=\epsilon_s+
2\epsilon_t\{z_0^{-2}-z_0^{-3}\}=0\;,\label{z0}
\end{eqnarray}   

\noindent or in the Cardano's form:

\begin{eqnarray}
&&z_0^3+p_0z_0+q=0; \;\;\mbox{where:}\;\; \displaystyle
p_0=-q={2\epsilon_t}/{\epsilon_s}\label{cardano}\\
&&\mbox{and:}\;\;
\displaystyle\left({q}/{2}\right)^2+
\left({p_0}/{3}\right)^3\equiv
\left({\epsilon_t}/{\epsilon_s}\right)^2+
\left({2\epsilon_t}/{3\epsilon_s}\right)^3>0.
\label{inequal}
\end{eqnarray}

\noindent The inequality in Eq. \ref{inequal} proves that there is a
unique real root of the cubic equation Eq. \ref{cardano}. Solution $z_0$ depends
only on the dimesionless parameter $p_0\equiv 2\epsilon_t/\epsilon_s$. A 
numerical solution of Eq. \ref{cardano} at $1/p_0=0.3$ gives:

\begin{equation}
 z_0=0.829\,.
\label{solz0}
\end{equation}

\noindent
Substitution of the solution $z_0$ into Eq. \ref{fredimless2} gives 
desired value of $f_0$.

\subsection{Free energy including patch and adhesion to the walls}

The free energy expressions Eq. \ref{fren} and Eq. \ref{frent} do not account 
neither for the adhesion of some part of the upper monolayer to the wall of the 
pipette, nor for the existence of the patch, which serves as a reservoir of 
lipids for the lower monolayer, see Fig. 1. In order to include these features of 
the experimental setting (Sukharev et al., 1999; Hase et al., 1995) we write 
the total Helmholtz free energy of the membrane as follows:

\begin{eqnarray}
&&F=\displaystyle N_1(f_1-f_0)+N_2(f_2-f_0+E_{ad})-N_1(a_1-a_{01})T_1-
N_2(a_2-a_{02})T_2-\nonumber\\
&&\displaystyle(N_2-N_{02})a_0\sin{\Theta}\cdot(T_1+T_2)-\sum_{i=1,2}
\lambda_i(S_i(H)-N_ia_i).
\label{frentot}
\end{eqnarray}

Explain the right hand side (rhs) expression in Eq. \ref{frentot} term 
by term. The first two terms signify the free energies of lipids in the curved 
parts of the monolayers $1$ and $2$ correspondingly. 
The energy per lipid molecule 
$f_0$ is 
subtracted in order to allow for the free energy difference 
between the molecules which belong to the curved part and to either the patch, 
or to the part of the monolayer $2$ that is stuck to the wall of the pipette.
In the latter case, we allow also for the adhesion, so that the energy per 
molecule is lowered with respect to the patch by an amount of the adhesion 
energy $E_{ad}$. The free
energies per lipid molecule $f_i$ in the i-th monolayer could be written using
definitions Eqs. \ref{fren}-\ref{frent} and dimensionless variables Eq. 
\ref{dimless} 
as follows: 

\begin{eqnarray}
&&f_1(z_1,h)\equiv
\epsilon_sz_1+\epsilon_t\left\{\displaystyle
({h^2}/{3})(7z_1^{-4}-4z_1^{-3})+
2h(z_1^{-2}-z_1^{-3})+(z_1^{-1}-1)^{2}\right\};
\label{fredimless1}\\
&&f_2(z_2,h)=f_1(z_2,-h)\label{fredimless0}.
\end{eqnarray}

The next two terms on the rhs of Eq. \ref{frentot} represent a mechanical work 
done by 
the tensions $T_{i=1,2}$ by stretching an area per molecule in the curved 
part of the i-th monolayer from the initial to the final value, $a_{0i}$ and
$a_i$ respectively. The initial areas $a_{0i}$, corresponding to the equilibrium
state at zero external pressure gradient, are calculated below together with 
the initial numbers $N_{0i}$ of the lipids in the curved parts of the monolayers. 
The fifth term on the rhs of
Eq. \ref{frentot} is introduced to allow for the mechanical work done by the 
perpendicular to the wall projections of the tensions during tearing off the 
2nd monolayer from the pipette's wall. Angle $\Theta$ is the wetting angle 
between the membrane and the wall:

\begin{eqnarray}
\sin{\Theta}=\{1-\cos^{2}{\Theta}\}^{1/2}=\displaystyle
\{1-\left(R/r\right)^2\}^{1/2}\equiv
\{1-\left({h}/{h_0}\right)^2\}^{1/2}\,,
\label{theta}
\end{eqnarray}

\noindent where notation defined in Eq. \ref{dimless} is used.
The factor $\sin{\Theta}\cdot(T_1+T_2)$ in the fifth term signifies
that the force acting on the 2nd monolayer perpendicular to the wall consists of
the $\sin{\Theta}\cdot T_2$ projection added to the projection of 
$\sin{\Theta}\cdot T_1$,
which is applied from the side of the 1st monolayer. The latter force equals, 
with an opposite sign, to 
the force acting perpendicular to the wall on the monolayer $1$ in accord with 
the Newton's 3rd law. Next,
the terms with the Lagrange 
multipliers $\lambda_{i=1,2}$ are introduced in order to select only the 
spherically shaped conformations of the membrane as its curvature $H$ changes 
together with the external pressure. In this case the external surface areas 
$S_1$ and $S_2$ of the monolayers and the numbers of the molecules in them are 
related by the following equations:

\begin{eqnarray}
&&S_1(H)=2\pi
(r-l_1)^2\left(1-\displaystyle\{1-{R^2}/{(r-l_1)^2}\}^{1/2}\right)=
N_1a_1\;
;
\label{area1}\\
&&S_2(H)=2\pi
(r+l_2)^2\left(1-\displaystyle\{1-{R^2}/{(r+l_2)^2}\}^{1/2}\right)=
N_2a_2\,.
\label{area2} 
\end{eqnarray}

\noindent Here the thicknesses $l_{i=1,2}$ of the i-th monolayer
are defined in Eqs. \ref{curvs1} and \ref{curvs2}. As long as the ratios 
are small:

\begin{eqnarray}
l_i/r\leq l_i/R\sim 10^{-3}\,,
\label{thin}
\end{eqnarray}

\noindent we shall neglect small corrections $l_i$ to the curvature radius $r$
in the expressions in Eqs. \ref{area1} and \ref{area2}, hence defining a 
unique surface area function $S(H)$:

\begin{eqnarray}
&&S(H)=2\pi r^2\left(1-\displaystyle\{1-{R^2}/{r^2}\}^{1/2}\right)\equiv
2\pi H^{-2}\left(1-\displaystyle\{1-(HR)^2\}^{1/2}\right)
\label{s(h)}
\end{eqnarray}

\noindent Then, using Eq. \ref{s(h)} and relations Eq. \ref{area1}, Eq. 
\ref{area2},
together with definitions Eq. \ref{dimless} we express the number of the lipids 
in the flat bilayer $N_0$ as follows:

\begin{eqnarray}
S_0=S(H\rightarrow 0)=\pi R^2=N_0a_0=N_0z_0{v}/{l_s}\,.
\label{s0}
\end{eqnarray}
\subsection{Equilibrium conformation under zero external pressure difference: 
$P=0$}
In the absence of pressure, the membrane is essentially flat, subjected to the
"resting" (Sukharev et al., 2001) tension arising from the membrane adhesion to 
the (glassy) surface of the pipette.
Thus, consider first a bilayer membrane in the equilibrium state in the pipette 
before an external pressure gradient is applied. Then, the tensions induced 
according to the Laplace's law Eq. \ref{laplace} by the external pressure 
difference $P$ are zero:

\begin{eqnarray}
T_1=T_2=0\,,
\label{p0}
\end{eqnarray}

\noindent and the free energy equals:

\begin{eqnarray}
F=\displaystyle N_{01}(f_1-f_0)+N_{02}(f_2-f_0+E_{ad})-\sum_{i=1,2}
\lambda_i(S(H)-N_{0i}a_{0i}).
\label{frentot0}
\end{eqnarray}

\noindent Corresponding equilibrium equations minimizing $F$ in Eq. (\ref{frentot0})
are derived in Appendix.
Using also relations Eq. \ref{area1} and Eq. \ref{area2}, we obtain the following 
closed form equation for the
unknown value of $H$ expressed in the dimensionless variable $h$ defined in 
Eq. \ref{dimless}: 

\begin{eqnarray}
&&\displaystyle\sum_{i=1,2} n_{i}{\partial f_i}/{\partial h}=-({1}/{N_0})
(\partial S(h)/\partial h)\sum_{i=1,2}{\partial f_i}/{\partial z_{i}}\,;
\label{equi0}\\
&&\mbox{where:  }\;\;\displaystyle n_i\equiv {N_i}/{N_0}=
2({z_0}/{z_i})(h_0/h)^2\left(
1-\{1-(h/h_0)^2\}^{1/2}\right)\,. 
\label{n_i}
\end{eqnarray}

\noindent The functions $f_i$ are defined in Eqs. \ref{fredimless1} and 
\ref{fredimless0}. 

\subsection{Semi-equilibrium conformation at finite external pressure 
difference $P$}

Soon after an application of a finite external pressure difference $P$ 
perpendicular 
to the plane of the membrane, i.e. during the times $\Delta t \leq t\ll \tau_s$, 
a semi-equilibrium conformation is achieved.
The free energy of the membrane in the semi-equilibrium state could be written
using Eq. \ref{frentot} as follows:

\begin{eqnarray}
F\rightarrow\displaystyle F-\lambda_3(N_1-N_2-N_{01}+N_{02})\,.
\label{frentot''}
\end{eqnarray}

\noindent Hence:

\begin{eqnarray}
&&F=\displaystyle N_1(f_1-f_0)+N_2(f_2-f_0+E_{ad})-
(N_2-N_{02})a_0\sin{\Theta}\cdot(T_1+T_2)-\nonumber\\
&&\displaystyle
\sum_{i=1,2}\left[N_i(a_i-a_{0i})T_i+\lambda_i(S_i(H)-N_ia_i)\right]-
\displaystyle\lambda_3(N_1-N_2-N_{01}+N_{02})\,.
\label{frentot'}
\end{eqnarray}

\noindent Here an extra Lagrange multiplier $\lambda_3$ is introduced to
reflect a simple physics. Namely, as explained in the Introduction, we consider
the semi-equilibrium state as existing during such a short time, which is not 
enough for a transfer of the lipid molecules from the patch to the curved 
part of the monolayer $1$. Hence, a change of the numbers of lipids 
$N_i(P)-N_{i0}$, in the curved parts of the monolayers $i=1,2$ relative to the 
initial, $P=0$ equilibrium conformation occurs only due to a tearing 
off the 2nd monolayer from the pipette's wall. In the latter case it is 
reasonable to assume that the numbers of lipids in both monolayers increase
simultaneously by the same amount (no interlayer slide):

\begin{eqnarray}
N_1-N_{01}=N_2-N_{02}\,.
\label{tear}
\end{eqnarray}

\noindent Here $N_{0i}$ are involved, which had to be found from the Eqs. 
\ref{N_01}-\ref{H}.
 Condition Eq. \ref{tear} is maintained by the term with $\lambda_3$ in Eq. 
\ref{frentot'}. As explained in the Introduction, due to small thickness
of the membrane Eq. \ref{thin} we assume equal lateral
tensions in the monolayers, which then could be found from the Laplace relation 
Eq. \ref{laplace}:

\begin{equation}
T_1=T_2=Pr/4\equiv T/2\,,
\label{laplace1}
\end{equation}

\noindent Here new parameter $T$ is introduced for convenience.
Corresponding equations for the minimum of Eq. \ref{frentot'} are presented 
in the Appendix.

\subsection{Relaxed conformation at finite external pressure 
difference $P$}

Equations for a complete (relaxed) equilibrium conformation of the membrane are
readily obtained, see Appendix, starting from the following expression for the 
free energy:

\begin{eqnarray}
&&F=\displaystyle N_1(f_1-f_0)+N_2(f_2-f_0+E_{ad})-
(N_2-N_{02})a_0T\sin{\Theta}-\nonumber\\
&&\displaystyle N_2(a_2-a_{02})T-\sum_{i=1,2}\lambda_i(S_i(H)-N_ia_i)\,.
\label{frentotc}
\end{eqnarray}

\noindent Equation \ref{frentotc} is obtained from
Eq. \ref{frentot'} by "relaxing" to zero the tension $T_1$ in the monolayer 
$1$,
and by dismissing a condition of a "dynamic cut off"  of the first monolayer 
from the reservoir of lipid molecules, i.e. from the patch. The latter 
condition was
represented in Eq. \ref{frentot'} by the term with the Lagrange multiplier 
$\lambda_3$. The lateral tension in the monolayer $2$ obeys the following 
Laplace relation:

\begin{equation}
T_2=Pr/2\equiv T\;\;\;\;; T_1=0\,.
\label{laplace2}
\end{equation}

\noindent
Corresponding equations for the minimum of Eq. \ref{frentotc} are presented 
in the Appendix.
\section{Discussion of numerical results}

\subsection{Initial equilibrium state}
Results obtained by the numerical solution of equations \ref{N_01}-\ref{H} 
are represented in Table 1. Despite the absence of the external
pressure gradient: $P=0$, the curvature $1/r$ proves to be small but finite:
$R/r\sim 10^{-3}$. This happens due to finite adhesion energy $E_{ad}$ of 
the monolayer $2$ to the wall of the pipette. In Table 1 $n_2$ 
is the number of molecules in the curved part of the monolayer $2$ normalized 
with the number $N_0$ in a free monolayer. Presented data indicate that $n_2$
decreases with an increase of the adhesion energy $E_{ad}$. The relative 
decrease of $n_2$ ranges from $0.8\%$ at $E_{ad}/\epsilon_t=0.03$ to $9\%$ at
$E_{ad}/\epsilon_t=0.3$. Simultaneously, the relative number of molecules $n_1$ 
($n_1=N_1/N_0$) in the monolayer $1$ remains practically constant and equals $1$.
The reason for the different behaviour of $n_1$ and $n_2$ is obvious.
More molecules from the curved part of the monolayer $2$ 
tend to stick to the wall as the adhesion energy $E_{ad}$ increases. On the 
other hand, though the monolayer $1$ leans to the wall following the shape of 
the monolayer $2$, it does not stick, hence it compensates for the "losses" of 
the number of molecules in its curved part by absorbing extra molecules from 
the patch.

Table 1 also contains the values of the areas $a_{1,2}$ per molecule in the 
monolayers, expressed in dimensionless units via the variables $z_1$, $z_2$ 
defined in Eq. \ref{dimless}. The range of $E_{ad}/\epsilon_t$ is chosen from 
the consideration of the stability of the membrane under stretching of its area,
which is known to have an upper threshold (Sukharev et al., 1999) of about 
$(a_i-a_0)/a_0\approx 4\%$. Remarkable is that in this way our theory predicts
reasonably well the value of the adhesion line tension $T_a$ of a lipid membrane 
to the glass wall known from the experiment (Opsahl and Webb, 1994).
Indeed, theoretically chosen ratio $E_{ad}/\epsilon_t=0.03$ in combination with
experimentally known absolute values $\epsilon_t\sim 4\cdot 10^{-13}erg$ and $a_0\sim 
60{\AA}^2$ leads to the estimate :

\begin{equation}
T_a\sim E_{ad}/a_0=2.5\,dyn/cm\,,
\label{adhesion}
\end{equation}

\noindent which falls right inside the experimentally measured interval 
$T_a=0.5\div 4\,dyn/cm$ (Opsahl and Webb, 1994). \\

Finally, the last line of Table 1 provides normalized values of the 
free energy of the membrane $F$, defined in Eq. \ref{frentot0}. It is 
interesting to notice that the values of $F/(N_0\epsilon_t)$  practically coincide 
with the corresponding values of the (normalized) adhesion energy per molecule 
$E_{ad}/\epsilon_t$, which are also indicated in Table 1. An examination of 
Eq. \ref{frentot0} leads to the conclusion that the 
coincidence is not accidental. Namely, while the reference of energy for the 
molecules in the monolayer $1$ is approximately $f_0$ per molecule, the one for 
the monolayer $2$ is $f_0-E_{ad}$. At zero external pressure $P=0$ the 
free energy per molecule in each monolayer approximately equals $f_0$, i.e. 
the flat membrane's value. Therefore, contribution of the monolayer $1$ to 
$F$ in Eq. \ref{frentot0} is nearly zero, and contribution of the monolayer 
$2$ is indeed $\approx E_{ad}N_0$, as long as $N_{02}\approx N_0$.

\subsection{Conformations at $P\neq 0$}

Numerical results were obtained by minimizing the free energy expressions 
of the memebrane in the pipette under a fixed external pressure. Expressions 
Eq. \ref{frentot'} and Eq. \ref{frentotc}  were used, describing correspondingly 
the semi- and complete equilibrium states of the membrane. The calculated 
theoretical dependences are presented in Figs. 2, and 3. Based on the 
experimental data (Sukharev et al., 1999; Hase et al., 1995; Sukharev et al., 
2001) we had chosen the following values for the two
dimensionless ratios controlling the main dependences:

\begin{equation}
 \epsilon_s/\epsilon_t=0.6;\;\;\;\;E_{ad}/\epsilon_t=0.03\,.
\label{choice}
\end{equation}

\noindent The first ratio in Eq. \ref{choice} sets the scale of the surface 
tension, while the second ratio fixes the scale of the adhesion energy of 
the membrane to the glassy wall of the pipette with respect to the energy of 
the entropic  repulsion between the hydrocarbon tails in the depth of
each monolayer.

As is shown in Fig. 2{\it{a}}, the free energy of the relaxed conformation is 
lower 
than that one in the semi-equilibrium state at any pressure difference $P$ 
measured in the dimensionless units introduced in Eq. \ref{dimp}:

\begin{equation}
 P\rightarrow p\equiv {Pv}/(2\epsilon_t)\,.
\label{dimp1}
\end{equation}

\noindent
The Fig. 2{\it{a}} justifies our conjecture made in the Introduction that a relaxation 
to zero of the lateral tension $T_1$ in the concave monolayer $1$ is concomitant 
with the decrease of the free energy $F$ of the bilayer membrane kept at a fixed 
external pressure difference. A slide of the monolayer $2$ is prevented by its 
finite adhesion enrgy to the wall of the pipette, i.e. $E_{ad}$.

Next, Fig. 2{\it{b}} provides an important evidence that the decrease of
the free energy and vanishing of $T_1$ happens by virtue of an increase of the
number of lipid molecules in the curved part of the monolayer $1$ at a constant 
pressure difference $P$. Indeed, 
$n_1^{conn}$ is manifestly greater than $n_1^{discon}$ at the same pressure.
Simultaneously, $n_2^{conn}$ is even slightly smaller than $n_2^{disc}$. The 
increase of $n_2$ under fast growth of the external pressure $P$ 
happens due to a tearing off the membrane from the glassy wall of the pipette. 
In the absence of the transfer of lipids from the patch, an "instant" increase 
of $n_1^{disc}$ with pressure in the semi-equilibrium state 
follows that one of $n_2^{disc}$ and has the same origin, i.e. tearing off the
membrane from the wall. But, after enough 
time had elapsed since the beginning of the pressure pulse, 
the $n_1$ shows an increase from $n_1^{disc}$ up to 
$n_1^{conn}$ at a fixed pressure due to a transfer of lipids from the patch.
The latter effect happens either via interlayer slide or/and via lipid diffusion,
as discussed in the Introduction. During this relaxation process the area per 
molecule in the monolayer $1$ decreases from 
$z_1^{disc}$ (in the dimensionless units) to practically that one in the free 
bilayer: $ z_1^{conn}\approx z_0$, see Fig. 3{\it{a}}. Simultaneously, the 
monolayer $2$ has to 
bear now the whole lateral tension: $T_2=T;\, T_1=0$ (previously shared with 
the monolayer $1$: $T_2=T_1=T/2$). The area per molecule in the monolayer $2$ 
stretches from $z_2^{disc}$ to $z_2^{conn}$. As long as an increase of this area 
relative to a free bilayer value exceeds $4\div 5\%$ the membrane becomes 
unstable. Therefore, we deduce from Fig. 3{\it{a}}, that the range of the external 
pressure differences in our model should be bound as follows: $0\leq P\leq 1.5$.

Results of our calculations of the membrane's curvature as a function of the 
pressure $P$ are presented in Fig. 2{\it{c}}. There is apparent effective 
softening of the membrane after a relaxation at constant $P$, i.e. 
$h_{conn}>h_{disc}$. 
Also, the curves exhibit tendency to a saturation of the curvature at hight 
enough pressures, in a qualitative accord with the experimental data (Sukharev 
et al., 1999).
Nevertheless, according to our results plotted in Fig. 3{\it{b}}, the lateral 
tension
$T_1$ continues to grow smoothly as a function of pressure $P$ even in the 
region of (near) saturation of the curvature.

Finally, we mention here, that separate consideration of the semi- and
complete equilibrium states discussed above, is justified by the fact that
the characteristic stress relaxation time: $\tau_{s,d}\sim 0.1\div 1 sec$
differs by 4$\div$5 orders of magnitude from the characteristic
time $\Delta t$ necessary for a propagation of the bending
deformations over a membrane with the linear dimensions $R\sim
1\mu m$:

\begin{eqnarray}
 \displaystyle \Delta t\sim R^{2}
 \left({\rho l_0}/{\epsilon_t}\right)^{1/2}
 \sim 10^{-8}cm^{2}\left({1g/cm^{3}
 \cdot 10^{-7}cm}/10^{-13}erg\right)^{1/2}
 \sim 10^{-5}sec\;.
\label{deltat}
\end{eqnarray}

\noindent Here we used a well known formula for the frequency of
the bending waves in a thin plate (Landau and Lifshitz, 1980):

\begin{eqnarray} \omega_b=q^2\left\{{k_B}/({2l_0\rho})\right\}^{1/2},
\label{omegab}
\end{eqnarray} 

\noindent where $k_B$ is the bending modulus, $q$ is a wave-vector, 
$\rho$ is the mass density of the membrane, and $2l_0$ is its thickness. 
Actually, the latter is dynamically increased by a drag of the adjacent layers 
of the liquid in which the membrane is immersed.
In Eq. \ref{deltat} we use $\epsilon_t$ as an estimate of $k_B$ involved in 
Eq. \ref{omegab} allowing for the role played by $\epsilon_t$ as a bending 
coefficient. This is justified by a direct comparison of the
free energy expressions Eqs. \ref{ftis1}, \ref{ftis2} and the
definition of $\epsilon_t$ given in Eq. \ref{dimless}.

\section{Summary}
Dynamics and characteristic times of bilayer membrane with MscL channels in the 
patch-clamp experimental setting are studied theoretically. The different characteristic 
times range from $10^{-5}\div 10^{-4}\,sec$ to several seconds. The shortest time 
corresponds to stress propagation along the micrometer-sized membrane bent by external
pressure gradient. The external pressure produces lateral tension in the bilayer, which
opens mechanosensitive channels (MscL). We show that lateral tension in lipid bilayer
does not remain constant under constant pressure gradient applied to the membrane.
Interlayer slide leads to redistribution of lateral stress within $0.01\div 0.1 sec$ after
application of the pressure step. Relaxation of lateral stress in the lower monolayer
triggers closing back of MscL channels ("adaptation"), observed experimentally, see companion 
paper. The longest time ($\sim 1\, sec$) arises from the free energy barrier for the transition 
from the open to the closed MscL conformation.

We have analysed the influence of bilayer hydrophobic thickness on the hydrophobic
mismatch energy and MscL thermodynamics. Thick bilayer hinders MscL opening, while thin 
bilayer favours the open conformation of the channels. Our estimates of the mismatch energy
give values in the range $2\div 20 k_BT_0$, where $k_BT_0$ is the thermal energy at room 
temperature $T_0$. This estimate is comparable with the free energies in the MscL 
elastic model (Sukharev et al., 1999) and (Sukharev and Markin, 2001). Depending on the 
bilayer thickness, the hydrophobic mismatch favours either closed or open MscL conformation,
as outlined in Section II. This may explain why MscL adaptation phenomenon, being present in 
the PC-20 lipid bilayers, is absent in the PC-18 bilayers (see companion paper).

We have studied the free energy functional of the membrane modeling the patch-clamp
experimental setting.  Minimizing the free energy functional, we have found parameters
of the membrane conformation as function of pressure gradient for two (semi- and complete)
equilibrium states. The semi-equilibrium state is an instantaneous mechanical equilibrium 
state achieved after application of external pressure step to the membrane. Following it,
a transfer of lipid molecules along the lower monolayer (see Fig. 1) takes place. It leads 
to the leveling of the finite lateral tension in the strained part of the lower monolayer 
with zero tension existing in the outer patch. As a result the free energy reaches its true 
minimum and the membrane achieves the complete equilibrium state. 
The upper monolayer is fixed due to lipid adhesion to the pipette glass wall. Our free 
energy functional of the membrane includes this adhesion energy. Theoretically predicted 
value of the adhesion line tension in our model is in a good agreement with experimental 
data (Opsahl and Webb, 1994).

The transfer of lipids in the lower monolayer may occur either via interlayer slide or/and 
via diffusion of lipids into the stretched region of the lower monolayer from the outer
patch (see Fig. 1), 
see also (Baoukina and Mukhin, 2003) and (Mukhin and Baoukina, 2002. Lipids diffusion 
mechanism of stress relaxation in a bilayer fluid membrane under pressure. cond-mat/0206099). 
Characteristic time of the individual lipid diffusion processes is by one order of magnitude 
longer than the time of the interlayer slide (collective motion of the lipids constituting the 
lower monolayer). Nevertheless, this times may become comparable at high enough density of 
MscL's, which then would pin the monolayer against slide.

The above explanation of the slide-triggered closing of a MscL leads to a proposal 
(Th. Schmidt, Leiden University, personal communication, 2002)
of a possible experimental check up of the theory. Namely, suppose that the 
membrane with MscL has been turned over, so that the cytoplasmic half of the 
channel belongs now to the monolayer $2$, which in turn stucks to the wall of 
the pipette.
Then, the opening of the channel would be controlled by the lateral tension 
$T_2$, which does not vanish (it even increases) after the relaxation of the 
membrane under a constant pressure. 
Hence, one would expect that adaptation of MscL after an application of the opening 
pressure would be less probable in this case.

Finally, we mention some possible future improvements of the
theory. These could be made by lifting the simplifying
restrictions of the homogeneity of the lateral stresses (tensions)
across the thickness of each monolayer. Besides, one may abandon the
approximation of a position-independent area per molecule on the 
membrane's surfaces. As a consequence, in a more elaborate
scheme of derivations one would not restrict himself to
the spherical symmetry of the membrane's shape. Nevertheless, we
believe that these improvements would not deny the main physical
concepts considered in this paper.

\vspace{5mm}
{\small{The authors are grateful to S.I. Sukharev for the introduction in
the problem and for numerous enlightening discussions during the
work. Useful discussions with Jan Zaanen and comments by R.F.
Bruinsma and Th. Schmidt are highly acknowledged.}} 

\newpage
\noindent
\begin{center}{REFERENCES}\end{center}
\vspace{3mm}

\noindent
Baoukina S.V. and S.I. Mukhin. 2003. Inter-layer slide mechanism of stress 
\par relaxation in bilayer fluid membrane under pressure. {\it{Biophys. J.}} 84: 1129a.\\
Ben-Shaul A. 1995. Molecular theory of chain packing, elasticity and 
\par lipid-protein interaction in lipid bilayers. {\it{In}} Structure and Dynamics of
\par membranes. Elsevier Science., 359-401\\
Betanzos Monica, Chien-Sung Chiang, H. Robert Guy and Sergei Sukharev. 2002. 
\par A large iris-like expansion of a mechanosensitive channel protein induced by 
\par membrane tension. {\it{Nature Structural Biology}} 9: 704-710\\
Cantor R.S. 1999. Lipid composition and the lateral pressure profile in bilayers. 
\par{\it{Biophys. J.}} 76: 2625-2639\\
Doyle D.A., J.M. Cabral, R.A. Pfuetzner, A. Kuo, J.M. Gulbis, S.L. Cohen,
\par B.T. Chait,  R. MacKinnon. 1998. The structure of the potassium channel: 
\par molecular basis of K+ conduction and selectivity. {\it{Science.}} 280:69-77\\
Fournier J.-B. 1999. Microscopic membrane elasticity and interactions among 
\par membrane inclusions: interplay between the shape, dilation, tilt and 
\par tilt-difference modes. {\it{Eur. Phys. J.}} 11:261-272\\
Gullingsrud J., K. Schulten. 2003. Gating of MscL studied by steered molecular 
\par dynamics. {\it{Biophys. J. Supplement}} 84: 21a\\
Hamill O.P. and B. Martinac. 2001. Molecular basis of mechanotransduction in 
\par living cells. {\it{Physiol. Rev.}} 81: 685-740\\
Hase C.C., A.C. Le Dain, B. Martinac. 1995. Purification and functional   
\par reconstitution of the recombinant large mechanosensitive ion channel of    
\par Escherichia coli. {\it{J. Biol. Chem.}} 270:18329-18334\\
Jian-Guo H.,O.Y. Zhong-Can. 1993. Shape equations of the axisymmetric
\par vesicles. {\it{Phys. Rev. E}} 47: 461-467\\
Landau L.D., E.M. Lifshitz 1980. Theory of Elasticity. V.7. Theoretical Physics. 
\par Pergamon Press, New York\\                                                        
Liu Q.H., H.G.  Zhou, J.X. Liu, O.Y. Zhong-Can. 1999. Spheres and prolate and      
\par oblate ellipsoids from an analytical solution of spontaneous curvature fluid       
\par membrane model. {\it{Phys. Rev. E}} 60:3227-3233\\
Ming Zhou, J.H. Morais-Cabral, S. Mann, R. MacKinnon. 2001. Potassium
\par channel receptor site for the inactivation gate and quaternary amine
\par inhibitors. {\it{Nature.}} 411:657-661\\
Opsahl L.R., W.W. Webb. 1994. Lipid-glass adhesion in giga-sealed 
\par patch-clamped membranes. {\it{Biophys. J.}} 66: 75-79\\
Sachs F. and C.E. Morris. 1998. Mechanosensitive ion channels in nonspecialized 
\par cells. {\it{Revs. Phyiol. Biocehm. Pharmacol.}} 132: 1-77\\
Safran S.A. 1994. Statistical thermodynamics of surfaces, interfaces and      
\par membranes. {\it{In}} Frontiers in Physics, Vol. 90. Perseus Pr. Publisher\\
Sonnleitner A., G.J. Schultz, and Th. Schmidt. 1999. Free Brownian motion of   
\par individual lipid molecules in biomembranes. {\it{Biophys. J.}} 77: 2638-2642\\
Sukharev S.I., W.J. Sigurdson, C. Kung, F. Sachs. 1999. Energetic and spatial    
\par parameters for gating of the bacteria large conductance mechanosensitive         
\par channel. {\it{J. Gen. Physiol.}} 113:525-539\\
Sukharev S., M. Betanzos, C.S. Chiang, H.R. Guy. 2001. The gating mechanism 
\par of the large mechanosensitive channel MscL. {\it{Nature.}} 409:720-724\\
Sukharev S.I. and V.S. Markin. 2001. Kinetic model of the bacterial large 
\par conductance mechanosensitive channel. {\it{Biological membranes}} 18:440-445\\
Xiang T.X., B.D. Anderson. 1994. Molecular distributions in interphases:   
\par statistical mechanical theory combined with molecular dynamics simulation   
\par of a model lipid bilayer. {\it{Biophys. J.}} 66:561-573\\ 
Zhong-Can O.Y., W. Helfrich. 1987. Instability and deformation of a spherical  
\par vesicle by pressure.{\it{Phys. Rev. Lett.}} 59:2486-2488\\
Zhong-Can O.Y., W. Helfrich. 1989. Bending energy of vesicle membranes.     
\par{\it{Phys. Rev. A}} 39:5280-5288                              
\newpage

{\appendix
\section{Equilibrium equations}                                                        
\subsection{Equilibrium conformation, $P=0$}

Considering $N_{0i}$, $a_{0i}$ and $H$ as the five independent 
variables,
we minimize $F$ in Eq. (\ref{frentot0}) with respect to all of them, and thus 
obtain the following equilibrium equations: 

\begin{eqnarray}
&&\displaystyle\partial F/{\partial N_{01}}=0:\;\;\{f_1-f_0\}/{a_{01}}
=-\lambda_1\,; \label{N_01}\\
&&\displaystyle\partial F/{\partial N_{02}}=0:\;\;\{f_2-f_0+
E_{ad}\}/{a_{02}}
=-\lambda_2\,; \label{N_02}\\
&&\displaystyle\partial F/{\partial a_{0i}}=0:\;\;\displaystyle
{\partial f_i}/{\partial a_{0i}}=-\lambda_i\;, \mbox{$i=1,2$}\,;
\label{a_0i}\\
&&\displaystyle{\partial F}/{\partial H}=0:\;\;\displaystyle 
\sum_{i=1,2}N_{0i}{\partial f_i}/{\partial H}=(\lambda_1+\lambda_2)
{\partial S(H)}/{\partial H}\,.\label{H}
\end{eqnarray}

\noindent One can easily see that Eqs. \ref{N_01} and \ref{N_02} permit
exclusion of the Lagrange multipliers $\lambda_i$ from the equations 
\ref{a_0i} and \ref{H}. Then, solving Eq. \ref{a_0i} for $i=1,2$ we find
dependences $a_{0i}(H)$, and substitute them into Eq. \ref{H}. In accord with
the discussion at the beginning of the Section III, after Eq. (\ref{frentot}), 
the Lagrange multiplier 
$\lambda_2$ in Eq. \ref{a_0i} plays a role of the "resting" tension in the layer 
$2$ due to a finite adhesion energy $E_{ad}$ of the layer to the glassy wall of 
the pipette.

\subsection{Semi-equilibrium conformation, $P\neq 0$}

In order to find the minimum of (\ref{frentot'}) we consider 
$N_{0i}$, $a_{0i}$ and $H$ as the five independent 
variables. Hence,
we minimize $F$ with respect to all of them, and thus obtain the following 
equilibrium equations: 

\begin{mathletters}
\begin{eqnarray}
\displaystyle{\partial F}/{\partial N_{1}}=0:&&\;\;\;f_1-f_0-({T}/{2})
(a_1-a_{01})-({T}/{2})a_0\sin{\Theta}=\lambda_3-\lambda_1a_1; \label{N_1d}\\
\displaystyle{\partial F}/{\partial N_2}=0:&&\;\;\;f_2-f_0+
E_{ad}-({T}/{2})
(a_2-a_{02})-({T}/{2})a_0\sin{\Theta}=\nonumber\\
&&-\lambda_3-\lambda_2a_2\,; \label{N_2d}\\
\displaystyle{\partial F}/{\partial a_{i}}=0:&&\;\;\displaystyle
{\partial f_i}/{\partial a_{i}}-{T}/{2}=-\lambda_i\;, \mbox{$i=1,2$}\,;
\label{a_id}\\
\displaystyle{\partial F}/{\partial H}=0:&&\;\;\displaystyle 
\sum_{i=1,2}N_{i}\left[{\partial f_i}/{\partial H}-
\partial (T/2)/\partial H(a_i-a_{0i})\right]-\nonumber\\
&&a_0(N_2-N_{02}){\partial(T\sin{\Theta})}/{\partial H}=
\displaystyle(\lambda_1+\lambda_2)
{\partial S(H)}/{\partial H}\,.\label{Hd}
\end{eqnarray}
\end{mathletters}

\noindent The system of equations Eqs. \ref{N_1d}-\ref{Hd} 
is solved numerically under condition Eq. \ref{tear},
after making a transformation to the dimensionless variables Eq. \ref{dimless} 
and
after normalization of the numbers $N_i$ by $N_0$ in accord with Eq. \ref{n_i}.

\subsection{Equilibrium (relaxed) conformation, $P\neq 0$}

Equations minimizing the free energy $F$ in Eq. \ref{frentotc}
take the form:

\begin{mathletters}
\begin{eqnarray}
\displaystyle{\partial F}/{\partial N_{1}}=0:&&\;\;\;f_1-f_0=
-\lambda_1a_1\,; \label{N_1c}\\
\displaystyle{\partial F}/{\partial N_2}=0:&&\;\;\;f_2-f_0+
E_{ad}-T(a_2-a_{02})-Ta_0\sin{\Theta}=-\lambda_2a_2\,; \label{N_2c}\\
\displaystyle{\partial F}/{\partial a_{1}}=0:&&\;\;\displaystyle
{\partial f_1}/{\partial a_{1}}=-\lambda_1\,;\label{a_1c}\\
\displaystyle{\partial F}/{\partial a_{2}}=0:&&\;\;\displaystyle
{\partial f_2}/{\partial a_{2}}-T=-\lambda_2\,;
\label{a_2c}\\
\displaystyle{\partial F}/{\partial H}=0:&&\;\;\displaystyle 
\sum_{i=1,2}N_{i}{\partial f_i}/{\partial H}-
N_2({\partial T}/{\partial H})(a_2-a_{02})-\nonumber\\
&&a_0(N_2-N_{02})
{\partial(T\sin{\Theta})}/{\partial H}=
\displaystyle(\lambda_1+\lambda_2)
{\partial S(H)}/{\partial H}\,.\label{Hc}
\end{eqnarray}
\end{mathletters}

\noindent The system of equations Eqs. \ref{N_1c}-\ref{Hc} is solved 
numerically after making a transformation to the dimensionless variables 
Eq. \ref{dimless} and Eq. \ref{n_i}.

\section{Elastic model of MscL channel}                                                        
\subsection{Phenomenological theory}
We consider a simple phenomenological model of MscL channel (Sukharev, Markin, 2001). 
The channel is presented as elastic cylinder; it's stiffness is different in closed and open conformations.
The free energy of the closed channel,$F_C$, is written as:

\begin{eqnarray}
F_C=\frac{B_C}{2}(A-A_C)^2-T(A-A_C)\,,
\label{fc}
\end{eqnarray}

\noindent
here $B_C$ is elastic modulus of the channel in the closed conformation;
$T$ - tension transmitted through the lipid bilayer;
$A_C$ - is the cross-section area of the closed channel in the absence of tension 
in lipid bilayer;
$A$ - is the channel cross-section area at a given tension $T$.
Transition to the open conformation of the channel is associated with stretching the 
channel area, rearrangement of transmembrane domains (helices) (Sukharev et al., 2001),
(Gullingsrud  and Schulten, 2003) 
and requires additional work. The work of opening the channel is done by tension in lipid 
bilayer. The free energy of the channel in the open state, $F_O$, equals:

\begin{eqnarray}
F_O=\frac{B_O}{2}(A-A_O)^2+E_O-T(A-A_C)\,,
\label{fo}
\end{eqnarray}

\noindent
here $B_O$ - elastic modulus of the open channel ($B_O\gg B_C$);
$A_O$ - the area of open channel at zero tension ($A_O > A_C$);
$E_O$ - the (free) energy difference between open and closed conformations of the channel 
at zero tension in the bilayer.
The finite energy $E_O$ reflects the fact that the probabilities to find the channel in the 
open or closed states ($P_O$ or $P_C$) are not equal (at zero tension the MscL is closed). 
In general, the value of $E_O$ depends on many factors, e.g. the type of lipids in the 
bilayer, thermodynamic 
conditions of the surrounding solution, and also on the 
hydrophobic mismatch energy, $F_{LP}$ (Hamill and Martinac, 2001). The latter is different 
for open and 
closed conformations of the channel, because MscL opening is associated with decrease of it's
hydrophobic thickness (S. Sukharev, Maryland University, personal communication, 2003). 
For the given (open or closed) state of the MscL channel, $F_{LP}$ 
contributes to the total free energy of this conformation ($F_O$ or $F_C$) and thus defines 
the value of $E_O$.
At tension $T$ the minima of the parabolas $F_C$  and $F_O$ define the free energies of the 
closed and open states, and the corresponding areas, $A^T_C$ or, $A^T_O$ are the areas 
occupied by closed or open channel in equilibrium:

\begin{eqnarray}
&&A^T_C=\frac{T}{B}+A_C\,,\;A^T_O=\frac{T}{B}+A_O\,,\\
&&F_C(A^T_C)=-\frac{T^2}{2B_C}\,,\;F_O(A^T_O)=-\frac{T^2}{2B_O}-T(A_O-A_C)+E_O\,.
\label{aco}
\end{eqnarray}

\noindent
The equilibrium distribution of open/closed channels,$n_O/n_C$ , at a given tension 
in membrane depends on the energy difference between open and closed conformations 
of the channel, $F_O-F_C$ :

\begin{eqnarray}
\frac{n_O}{n_C}=\frac{P_O}{P_C}\sim\exp{\{-(F_O-F_C)/k_BT_0\}}\,,
\label{nratio}
\end{eqnarray}

\noindent
here $n_O$ and $n_C$ are fractions of open and closed channels, respectively;
$k_B$ - Boltzmann constant; $T$ - temperature.
The intersection point of the energy parabolas $F_C(A)$ and $F_O(A)$ is the barrier energy 
for transition between closed and open states, $F_{bar}(A_b)$:

\begin{eqnarray}
F_C(A_b)=F_O(A_b)=F_{bar}(A_b)\,.
\label{ab}
\end{eqnarray}

\noindent
By comparing (\ref{fc}) and (\ref{fo}), one finds that the area $A_b$ ($A_C<A_b<A_O$) does 
not depend on the tension, $T$. 
The activation energies for closing and opening of the channel, $E_{act,c}$ and $E_{act,o}$:

\begin{eqnarray}
E_{act,c}=F_{bar}(A_b)-F_O(A^T_O)\,,\; E_{act,o}=F_{bar}(A_b)-F_C(A^T_C)
\label{eco}
\end{eqnarray}

\noindent
define the kinetics of transitions between closed and open states. 
The activation barriers for closing and opening of the channel equal:

\begin{eqnarray}
&&E_{act,c}=F_{bar}(A_b)-F_O(A^T_O)=\frac{B_O}{2}(A_O-A_b)^2+T(A_O-A_b)+
\frac{T^2}{2B_O}\,;
\label{barc}\\
&& E_{act,o}(T)=\frac{B_0}{2}(A_b-A_C)^2-T(A_b-A_C)+\frac{T^2}{2B_C}\,.
\label{baro}
\end{eqnarray}

\noindent Solving Eq. (\ref{ab}) with respect to $A_b$ and substituting the result 
into Eq. (\ref{barc}), one finds dependence $E_{act,c}$ on $E_0$, shown in Fig. 5 for
the different values of the tension $T$. As could be discerned already in Fig. 4,
$E_{act,c}$ decreases with the increase of $E_0$.

The equilibrium 
distribution of closed and open channels (\ref{nratio}) is reached within the characteristic 
time $\tau_{eqv}$:

\begin{equation}
\tau_{eqv}=(k_O+k_C)^{-1}\,,
\label{taue}
\end{equation}

\noindent
here $k_O$ and $k_C$ are the rates of opening and closing of the channels. 
The value of the rate constant for the open to closed transition, $k_C$ , is the key 
factor for the channel closing induced by interlayer slide (as $n_C\rightarrow 1$ ):

\begin{equation}
k_C=\tilde{k}\exp{\{-E_{act,c}/k_BT_0\}}\,.
\label{kc}
\end{equation}

\subsection{The energy of hydrophobic mismatch}
The mismatch of hydrophobic thickness of the protein and that of the surrounding lipid 
bilayer makes lipids adjust to the protein length. This induces local deformation of 
lipid molecules. The deformation profile, $d(r)$, can be presented in a simple exponential 
form (Ben-Shaul, 1995):

\begin{equation}
d(r)=d_0+(d_p-d_0)\exp{\{-(r-r_p)/r_0\}}\,,
\label{deform}
\end{equation}

\noindent here:
$r$ -distance from the protein center in the membrane plane;
$r_p$ -radius of the protein (protein is modeled as a cylinder);
$d_0$ - the (equilibrium) thickness of the bilayer hydrophobic core in 
the absence of protein;
$d_p$ - is the thickness of protein's hydrophobic region;
$r_0$  - the characteristic scale of the deformation.
For the sake of simplicity let us consider a flat bilayer. The energy 
(per unit area) 
of elastic deformation of the bilayer by a cylindrical protein, $f_{LP}$, 
to the lowest order (neglecting tilt of lipid chains) can be written as 
(Fournier, 1999):

\begin{equation}
f_{LP}=\frac{K_d}{2}(d-d_0)^2+\frac{K_g}{2}(\nabla d)^2\,,
\label{flp}
\end{equation}

\noindent
here $K_d$ is the dilation modulus, $K_d\sim 2K_A/(2l_0)^2$;
$K_g$ - is the modulus, characterizing the energy cost of producing a gradient of 
bilayer thickness (it includes the energy of increasing the area of chain-water 
interface (Fournier, 1999)).\\
We consider the concentration of the proteins to be small and calculate the 
hydrophobic mismatch energy, $F_{LP}$, per channel as follows:

\begin{equation}
F_{LP}=\int^{\infty}_{r_p}f_{LP}2\pi rdr\,.
\label{FLP}
\end{equation}

\noindent
The result of the integration is:

\begin{equation}
F_{LP}=\displaystyle\frac{\pi}{4}(K_dr^2_0+K_g)
\left(\frac{2r_p}{r_0}+1\right)(d_p-d_0)^2\,.
\label{Flpres}
\end{equation}

\noindent
Assuming $K_d\sim 15\cdot 10^{14} erg/cm^4$ (Hamill and Martinac, 2001), 
$K_g\sim 35\,erg/cm^2$ (Fournier, 1999), $r_0 \approx 10\AA$ 
(Ben-Shaul, 1995), $r_p\approx 25 \AA$ (Sukharev et al., 1999), 
we obtain the following estimates for $F_{LP}$ for two different values of hydrophobic 
mismatch, $d_p-d_0$ :\\
1) for $|d_p-d_0|\sim 2\AA$, $F_{LP}\sim 2.3\, k_BT_0$ at room temperature $T_0$;\\
2) for $|d_p-d_0|\sim 6\AA$, $F_{LP}\sim 21\, k_BT_0$;\\
here $k_B$ is the Boltzmann constant, $T_0$ is temperature.
The value of $d_p-d_0$  depends on the type of lipids (i.e. the length of 
hydrocarbon chains) and on the state of the protein channel.  
}

\newpage

\begin{table}
\caption{Numerical results: initial equilibrium at $P=0$}
\begin{tabular}{lccc}
\tableline
$\displaystyle E_{ad}/\epsilon_t=0.03$&$\displaystyle E_{ad}/\epsilon_t=0.06$&
$\displaystyle E_{ad}/\epsilon_t=0.1$&$\displaystyle E_{ad}/\epsilon_t=0.3$\\
\tableline
$z_2/z_0=1.0077$&$z_2/z_0=1.0158$&$z_2/z_0=1.027$&$z_2/z_0=1.093$\\
$n_2=0.9923$&$n_2=0.9844$&$n_2=0.9737$&$n_2=0.9151$\\
$\displaystyle F/(N_0\epsilon_t)=0.03$&$\displaystyle F/(N_0\epsilon_t)=
0.0596$&$\displaystyle F/(N_0\epsilon_t)=0.0988$&
$\displaystyle F/(N_0\epsilon_t)=0.2878$\\ \tableline
$z_1=z_0=0.8291$,&  $n_1=1.000$ & for all values of 
& $\displaystyle E_{ad}/\epsilon_t$.\\ 
\end{tabular}
\end{table}

\newpage
\vspace{3mm}

\begin{center}
{\bf Figure Legends}
\end{center}

\vspace{3mm}

Fig. 1. Sketch of a membrane in the experimental patch-clamp settings. 
The numbering 1,2 of the monolayers within a bilayer membrane is indicated.

\vspace{5mm} Fig. 2{\it{a}}. Normalized free energy $F/{(\epsilon_t N_0)}$ in the 
semi-equilibrium
$F_{disc}$ and complete equilibrium $F_{conn}$ state {\it{versus}}
dimensionless pressure difference $p$ (see Eq. \ref{dimp1} ); 
$\epsilon_s/\epsilon_t=0.6$.

\vspace{5mm} Fig. 2{\it{b}}. Normalized numbers of lipid molecules in the curved parts
of the membrane monolayers $n_{i=1,2}$ in the semi-equilibrium
$n_i^{disc}$ and complete equilibrium $n_i^{conn}$ state (see Eq. \ref{n_i} ) 
{\it{versus}} dimensionless pressure difference $p$. Other
parameters are as in Fig. 2{\it{a}}. 

\vspace{5mm}  Fig. 2{\it{c}}. Normalized curvature of the bilayer membrane inside the 
pipette in the semi-equilibrium $h_{disc}/h_0$ and complete equilibrium 
$h_{conn}/h_0$ state {\it{versus}} dimensionless pressure difference $p$; 
$h/h_0\equiv R/r$, where $R$ and $r$ are the pipette 
and the curvature radii respectively. Other parameters are as in Fig. 2{\it{a}}. 

\vspace{5mm}  Fig. 3{\it{a}}. Normalized areas per lipid molecule at the external surfaces
of the monolayers inside the pipette in the semi-equilibrium $z_i^{disc}/z_0$ and 
complete equilibrium $z_i^{conn}/z_0$ state {\it{versus}} dimensionless pressure 
difference $p$. Other parameters are as in Fig. 2{\it{a}}.

\vspace{5mm}  Fig. 3{\it{b}}. {\it{Dots}}: Normalized lateral tension $T_1$ in the monolayer 
$1$ in the semi-equilibrium state {\it{versus}} dimensionless pressure difference 
$p$. {\it{Solid line}}: Normalized curvature $h_{disc}/h_0$ of the membrane 
inside the pipette in the semi-equilibrium state is taken from Fig. 2{\it{c}} 
for a convenience of comparison. Other parameters are as in Fig. 2{\it{a}}.
 
\vspace{5mm}  Fig. 4. Sketch of the free energy of MscL as function of the pore area $A$
in the elastic model (Sukharev and Markin, 2001). Solid lines - zero tension $T=0$.
The grey lines indicate free energy at finite tension $T$. The less intensive are the lines - 
the greater is the tension. 

\vspace{5mm} Fig. 5.
The activation energy (barrier) for channel closing, $E_{act,c}$ as a function of the free 
energy difference between open and closed channel conformations (at zero tension) $E_0$. 
Solid black line corresponds to zero tension $T=0$; the gray lines correspond to the tensions 
$T=6 dyn/cm$ and $T=12 dyn/cm$ (less intensive line is for the greater tension). The energies 
are expressed in units of $k_BT_0$. Other parameters: 
$B_C=0.22 k_BT_0;\; B_O=2.2 k_BT_0;\; A_C=13 nm^2;\; A_O=30 nm^2$ (Sukharev, Markin, 2001). 


\newpage
\begin{figure}
 \vbox to 4.0cm {\vss\hbox to -5.0cm
 {\hss\
       {\includegraphics{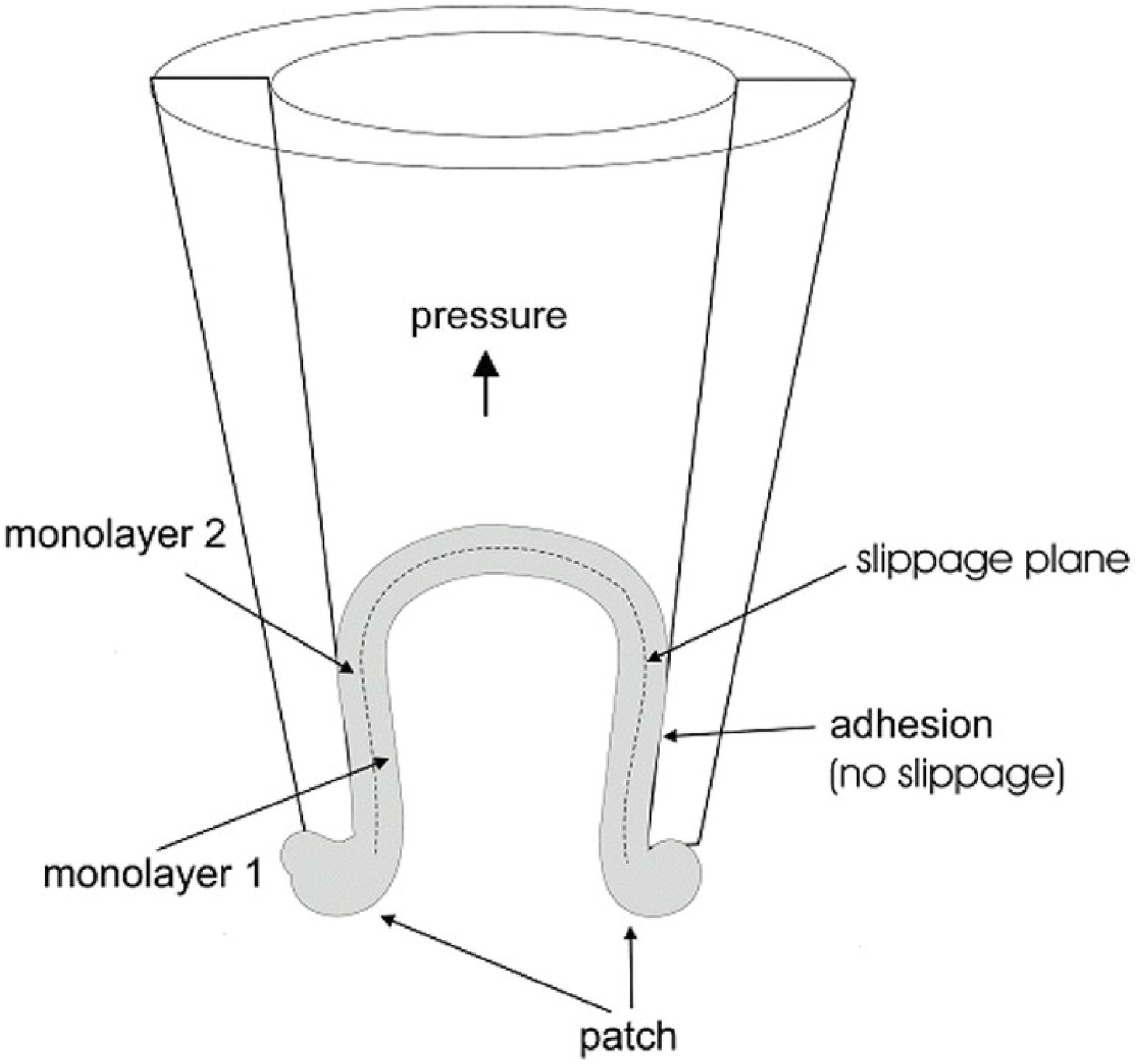}
       }
  \hss}
 }
\vspace{17cm} \caption{}  \label{patch}
\end{figure}

\newpage

\begin{figure}
 \vbox to 4.0cm {\vss\hbox to -5.0cm
 {\hss\
       {\includegraphics{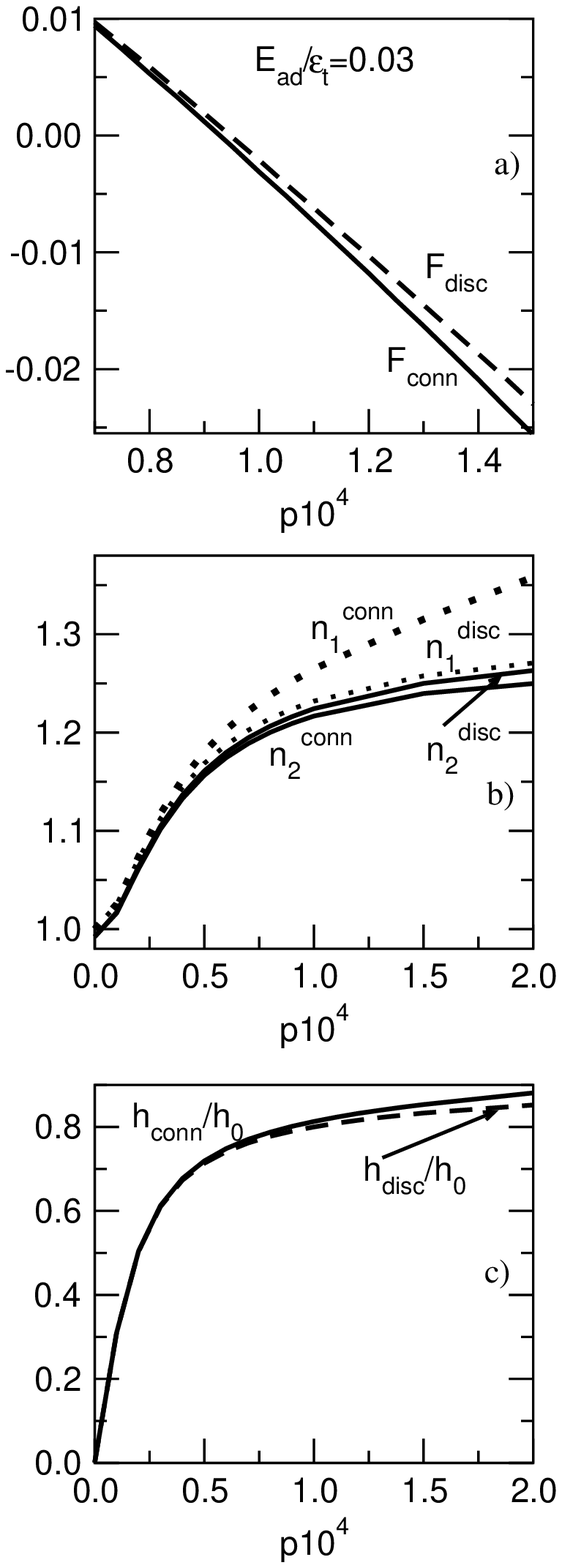}
       }
  \hss}
 }
\vspace{17.5cm}\caption{} \label{frens}
\end{figure}
\newpage

\begin{figure}
 \vbox to 4.0cm {\vss\hbox to -5.0cm
 {\hss\
       {\includegraphics{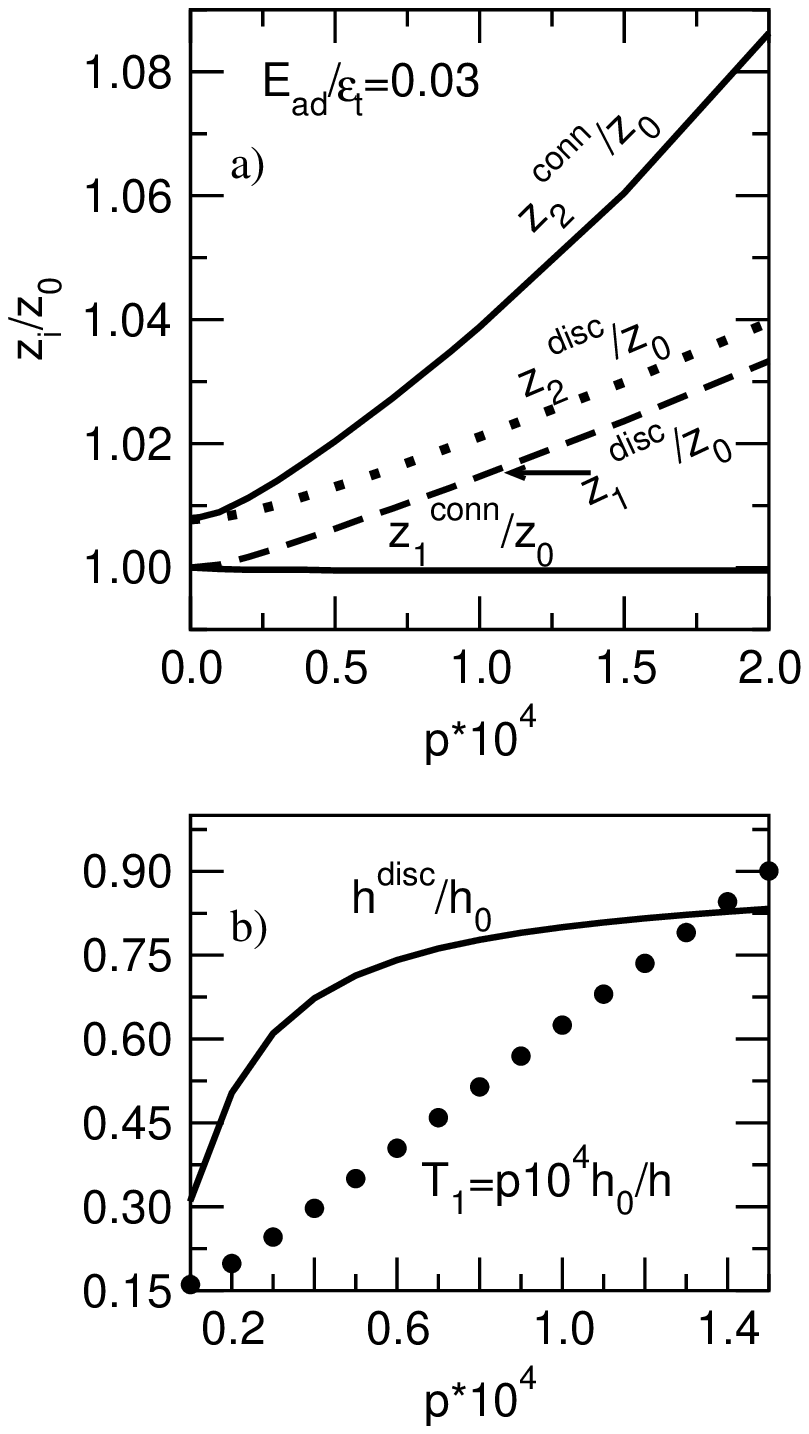}
       }
  \hss}
 }
\vspace{17.5cm}\caption{} \label{zis}
\end{figure}

\newpage

\begin{figure}
 \vbox to 4.0cm {\vss\hbox to -5.0cm
 {\hss\
       {\includegraphics{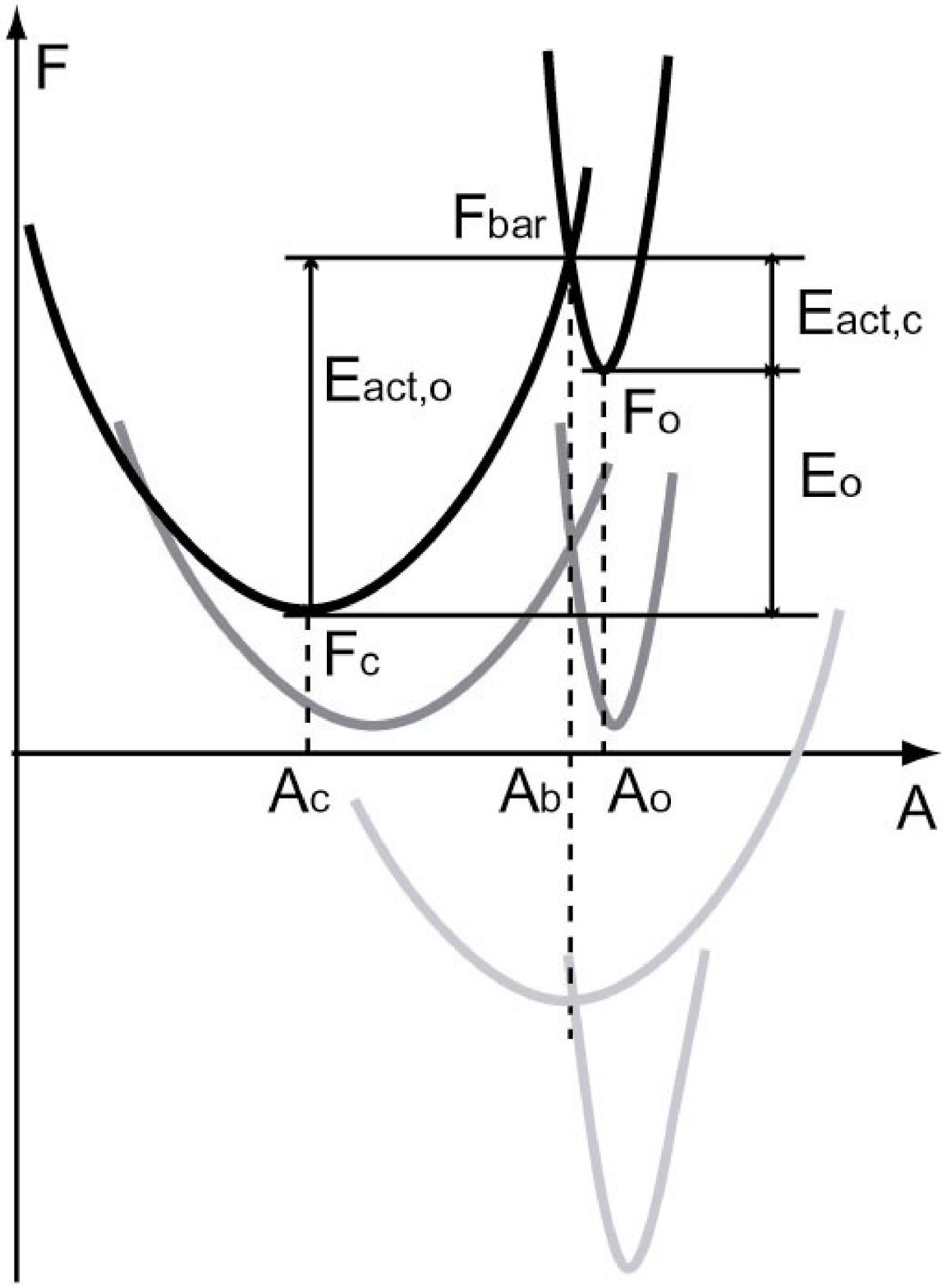}
       }
  \hss}
 }
\vspace{17.5cm}\caption{} \label{mscl}
\end{figure}

\newpage

\begin{figure}
 \vbox to 4.0cm {\vss\hbox to -5.0cm
 {\hss\
       {\includegraphics{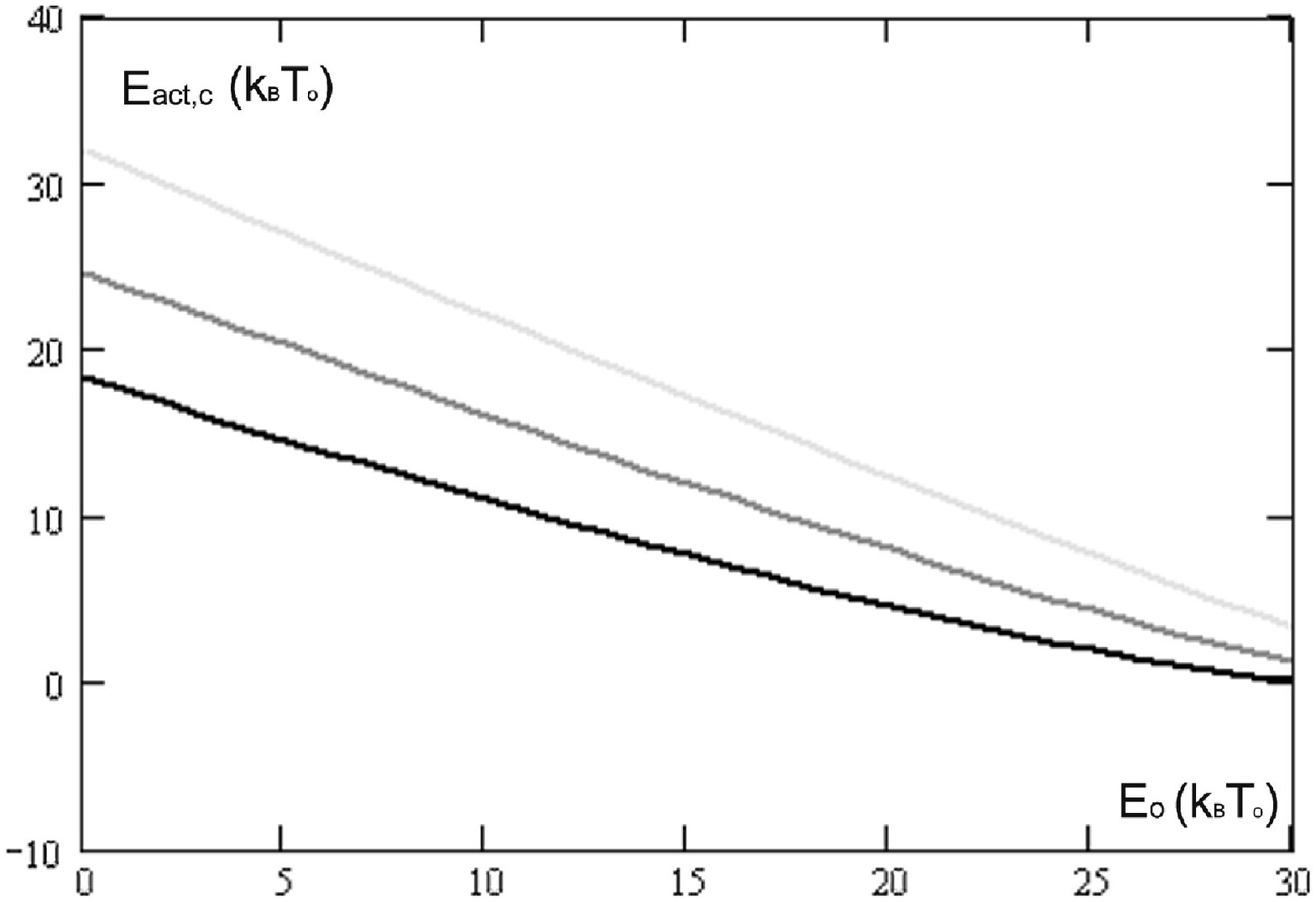}
       }
  \hss}
 }
\vspace{17.5cm}\caption{} \label{frens}
\end{figure}
\newpage

\end{document}